\documentclass[aps,prl,twocolumn,superscriptaddress,longbibliography]{revtex4-2}

\usepackage{amsmath,amssymb}
\usepackage{graphicx}
\usepackage{xcolor}
\usepackage{hyperref}
\usepackage{bm}
\usepackage{siunitx}
\usepackage{upgreek}
\usepackage[most]{tcolorbox}
\usepackage{afterpage}

\definecolor{pnasblue}{RGB}{0,48,135}
\definecolor{pnasbg}{RGB}{236,242,252}

\newtcolorbox{significancebox}{
  enhanced,
  colback=pnasbg,
  colframe=pnasblue,
  boxrule=1.5pt,
  arc=0pt,
  left=6pt, right=6pt, top=6pt, bottom=6pt,
  title={\bfseries\color{pnasblue}Significance},
  attach boxed title to top left={yshift=-2mm, xshift=4mm},
  boxed title style={colback=white, colframe=pnasblue, boxrule=1pt, arc=0pt,
    left=3pt, right=3pt, top=1pt, bottom=1pt},
  fonttitle=\small\sffamily,
  fontlower=\small
}

\usepackage{fancyhdr}

\begin{document}

\preprint{arXiv:XXXX.XXXXX [cond-mat.soft]}

\title{Short-range electrostatic screening in ionic liquids\\
as inferred by direct force measurements}

\author{Benjamin Cross$^*$}
\affiliation{Universit\'e Grenoble-Alpes, CNRS,
Laboratoire Interdisciplinaire de Physique,
Grenoble Cedex 9 38041, France}

\author{L\'eo Garcia}
\affiliation{Universit\'e Grenoble-Alpes, CNRS,
Laboratoire Interdisciplinaire de Physique,
Grenoble Cedex 9 38041, France}

\author{Elisabeth Charlaix}
\affiliation{Universit\'e Grenoble-Alpes, CNRS,
Laboratoire Interdisciplinaire de Physique,
Grenoble Cedex 9 38041, France}

\author{Patrick K\'ekicheff$^*$}
\affiliation{Institut Charles Sadron, Universit\'e de Strasbourg,
CNRS UPR22, Strasbourg Cedex 2 67034, France}

\date{\today}

\begin{abstract}
Previous experimental reports of long-range interactions in ionic
liquids (ILs) stand in contradiction with theoretical predictions
and numerical simulations. To provide insights into the literature
discrepancies regarding the experimental ranges of electrostatic
screening, claimed with orders of magnitude larger, the interactions
between pairs of mica and borosilicate surfaces confining ILs are
investigated by two complementary advanced Surface Force
Apparatuses. Regardless of differences in confinement geometries
(crossed-cylinders, sphere-flat), radii of curvature (cm--mm), and
measurement techniques (stepwise versus continuous approach), two
ever present force regimes are evidenced. At small surface
separations, oscillatory forces reflect IL structuration and
layering, while outside this gap, the interaction is monotonic
repulsive. In both regimes the spatial extent and force magnitude
depend critically on motion conditions, as demonstrated by
achieving velocities as low as 9~pm/s with equilibration times up
to 90~s. At large separations, fast surface displacements generate
long-range interactions (over tens of ion size) creating the
illusion of anomalous underscreening, whereas increasingly slow
ones shrink both magnitude and range of the repulsion with
decay-lengths converging ultimately to a screening length
consistent with Poisson--Boltzmann theory with finite ion sizes.
The transition from apparent long-range to short-range screening
unfolds over nearly two orders of magnitude in time, revealing slow
relaxation dynamics reminiscent of aging phenomena. These findings
definitely resolve a decade-old controversy on force measurements
and reveal rich out-of-equilibrium dynamics. The hydrodynamic
contribution to the net force is admittedly crucial to be reduced
especially when relaxations span decades in time, but approaching
thermodynamic equilibrium during measurements proves essential.
\end{abstract}


\maketitle

\newpage
\begin{tcolorbox}[
  enhanced, colback=pnasbg, colframe=pnasblue, boxrule=1.5pt,
  arc=0pt, left=6pt, right=6pt, top=5pt, bottom=5pt,
  title={\bfseries\small\sffamily\color{pnasblue}Significance},
  attach boxed title to top left={yshift=-2mm, xshift=4mm},
  boxed title style={colback=white, colframe=pnasblue, boxrule=1pt,
    arc=0pt, left=3pt, right=3pt, top=1pt, bottom=1pt}
]
{\small In electrolytes mobile ions screen electric fields all the more
they are concentrated. However, in the last decade, conflicting
experimental reports of long-range interactions in ionic liquids
(ILs) and concentrated electrolytes have puzzled theory and
numerical simulations. The discrepancy lies in nonequilibrium
effects as evidenced by our rigorous measurement protocols using
complementary advanced Surface Force Apparatuses. ILs screening
is genuinely short-range under conditions approaching
thermodynamic equilibrium. Their decay lengths ($\sim$0.5~nm) match
the theoretical predictions. Conversely the further from
equilibrium the larger interaction extents. These arise from
background drift rates with various relaxations spanning two orders
of magnitude in time scale. The IL behavior is reminiscent of slow
relaxation dynamics in glasses, thin polymer films, historical
paintings, and granular media.}
\end{tcolorbox}

{\fontsize{7}{8.5}\selectfont
\noindent\textbf{Author contributions:} B.C.\ and P.K.\ designed research; L.G., B.C., and P.K.\
performed the experiments; B.C.\ and P.K.\ analyzed data; B.C., P.K., and E.C.\ interpreted
the results; and P.K.\ and B.C.\ wrote the paper with the input of all authors.
All authors gave approval to the final version of the paper.

\noindent\textbf{Competing interests:} The authors declare no competing interest.

\noindent\textbf{Correspondence:} \href{mailto:benjamin.cross@univ-grenoble-alpes.fr}{benjamin.cross@univ-grenoble-alpes.fr};
\href{mailto:patrick.kekicheff@ics-cnrs.unistra.fr}{patrick.kekicheff@ics-cnrs.unistra.fr}

\noindent Published in \textit{PNAS} \textbf{123}(7), e2517939123 (2026).
\href{https://doi.org/10.1073/pnas.2517939123}{doi:10.1073/pnas.2517939123}

\noindent\copyright\ 2026 the Author(s). Distributed under
\href{https://creativecommons.org/licenses/by-nc-nd/4.0/}{CC~BY-NC-ND~4.0}.
}


\section*{Introduction}

Ionic liquids are molten salts comprising polyatomic organic or
inorganic ions that melt at temperatures below 100~$^\circ$C. Their
cations and anions are typically of large size (5 to 10 larger than
monoatomic ions), asymmetric shape with conformational flexibility,
polar, polarizable, often hydrophobic~\cite{armand2009,hayes2015}.
Unlike traditional salts, these anisotropic flexible and large
structures separate cation and anion centers, lowering the ionic
liquids (ILs) melting points to ambient temperature. The short-range
ion--ion interactions concur with the shape and the charge
distribution of ion molecules to unique IL properties. Their
nonflammability, low volatility, high ion conductivity, thermal and
chemical stability, enable numerous applications in bulk (catalysis,
electrochemistry) and at interfaces (lubrication, electrowetting,
energy storage)~\cite{fedorov2014}. In confined geometries the solid
interfaces may change the properties because of the IL interactions
with the substrate and of a layering IL structure formation, as
revealed both by molecular dynamics~\cite{lyndellbell2007,rotenberg2018,coles2020,zeman2021,reinertsen2024}
and experimental techniques sensitive to molecular arrangements at
interfaces, such as X-ray reflectivity~\cite{mezger2008}, atomic
force microscopy (AFM)~\cite{hayes2015,atkin2011,werzer2012,hjalmarsson2016,hjalmarsson2017},
or surface force apparatus
(SFA)~\cite{horn1988,perkin2010a,ueno2010,boumalham2010,perkin2011,perkin2012,smith2013a,smith2013b,espinosa2014,gebbie2015,cheng2015,smith2015,smith2016,lee2017,gebbie2017,comtet2017,garcia2018,lhermerout2018}.
These interface and confinement properties, which have only been
partially explored remain not fully understood.

First, spontaneous phase change from liquid to solid may occur when
the IL is squeezed between solid walls into films thinner than
typically 5 to 10 molecular diameters. Thus, some studies suggest
ILs behave partly as solids under confinement~\cite{comtet2017,chen2007},
while others disagree~\cite{ueno2010,garcia2018,gobel2009}. Second,
between confining walls, beyond the short-range ordered IL layering
regime perpendicular to the interfaces, most of the force
measurements did not reveal any other interactions either by
AFM~\cite{atkin2011,werzer2012,hjalmarsson2016,hjalmarsson2017} or
SFA~\cite{horn1988,perkin2010a,ueno2010,boumalham2010,perkin2011,perkin2012,smith2013a,smith2013b,cheng2015,smith2015}
techniques. Nevertheless, recent studies show interactions decaying
exponentially at large surface separation $D$. These observations
[mainly SFA~\cite{espinosa2014,gebbie2015,smith2015,smith2016,lee2017,gebbie2017,comtet2017,lhermerout2018};
one AFM study~\cite{hjalmarsson2017}] prompted comparisons with the
exponential repulsion operating at large distances in aqueous
electrolytes~\cite{pashley1984a,pashley1984b,pashley1984c,shubin1993,kekicheff1993,baimpos2014,gaddam2019,kumar2022}.
The main focus was on the decay-length, identified as the screening
length, $\lambda$, of the electrical double layer. In the limit of
low concentrations, the corresponding decay-length is the Debye
screening length, $\kappa_\mathrm{D}^{-1}$, that decreases with the
total ionic density as
\begin{equation}
  \kappa_\mathrm{D}^{-1} =
  \left(\frac{\varepsilon_0\varepsilon_\mathrm{r}k_\mathrm{B}T}
  {\sum_j\rho_j q_j^2}\right)^{1/2}
\end{equation}
where $k_\mathrm{B}$ designates the Boltzmann constant, $T$ the
temperature, $\varepsilon_0$ the vacuum permittivity, and
$\varepsilon_\mathrm{r}$ the relative permittivity (dielectric
constant) and where the sum runs over the charged species with
number density $\rho_j$ and charge $q_j$. In a 1:1 electrolyte of
bulk concentration $\rho_\mathrm{ion}$, the Debye length is
$\kappa_\mathrm{D}^{-1} = 1/\sqrt{8\pi L_\mathrm{B}\rho_\mathrm{ion}}$
with $L_\mathrm{B} = e^2/4\pi\varepsilon_0\varepsilon_\mathrm{r}k_\mathrm{B}T$
the Bjerrum length of the solvent ($\approx 0.7$~nm for water at
room temperature). When the screening length becomes comparable to
the ion diameter $a$ (bare or hydrated size), or to the mean
distance between charges, the continuum approximation for the charge
density becomes inappropriate and corrections to the Debye length
are expected~\cite{attard1993,janecek2009}. Thus, a few experimental
reports~\cite{smith2016,pashley1984a,baimpos2014,gaddam2019} revealed
large decay-lengths, $\lambda$, in concentrated aqueous electrolytes,
while others did not~\cite{pashley1984b,pashley1984c,shubin1993,kekicheff1993,kumar2022}.
Furthermore, $\lambda$ appeared to increase with ionic concentration,
strongly exceeding expected $\kappa_\mathrm{D}^{-1}$
values~\cite{smith2016,pashley1984a,baimpos2014,gaddam2019}. These
anomalously large screening lengths in concentrated aqueous
electrolytes were reunited to the other few similar IL observations,
and this set, so-called ``underscreening'' behavior, was claimed to
follow an empirical scaling relation:
\begin{equation}
  \frac{\lambda}{\kappa_\mathrm{D}^{-1}} \propto
  \left(a\kappa_\mathrm{D}\right)^\nu
  \label{eq:underscreening}
\end{equation}
with $\nu = 0$ for $\kappa_\mathrm{D}^{-1} > a$, and $\nu = 3$ for
$\kappa_\mathrm{D}^{-1} < a$~\cite{lee2017}.

The concept of constraint due to the ion size is evidently fruitful,
especially in the case of large ions. Several theories taking into
account the ions sizes have been suggested for
long~\cite{attard1993,janecek2009}, and the unexpected large
underscreening behavior has motivated several recent theoretical
approaches (see for instance refs.~\cite{rotenberg2018,coles2020,zeman2021,adar2019}
and references therein). However, despite ingenious interpretations
of Eq.~\eqref{eq:underscreening}, no theoretical model has
reproduced this scaling relation. Thus, the exponent $\nu$ is
calculated to be around 2 rather than 3 in ref.~\cite{adar2019}.
Such prediction comes also along similar conclusions established by
numerical simulations: ``in a high density regime the ratio
$\lambda/\kappa_\mathrm{D}^{-1}$ increases according to a power
law, in qualitative agreement with experimental measurements, albeit
at a much slower rate''~\cite{rotenberg2018}. There, the exponent
$\nu \sim 3/2$, well below the reported 3, does evidence an
underscreening behavior, that nevertheless does not lead to the
suggested extent of long-range exponential repulsions with large
screening lengths.

Whether these discrepancies result from theoretical limitations,
simulation imperfections, or experimental shortcomings and
inadequacies remains unclear. It is indeed a challenge to grasp
complex processes when competing ion adsorption on charged surfaces,
ion layering, cluster formation, and other electrical double layer
phenomena (dynamic and structural behaviors) are involved.
Elaboration of such a complex and multiparametric model can be
based only on a versatile collection of information obtained by
reliable sets of experimental data. Despite extensive research in
concentrated aqueous electrolytes and in ILs, experimental
force-distance profiles remain contradictory, both in their range
and in their magnitude. Furthermore, for the observations claiming
underscreening behavior, the reported screening length values exceed
theoretical predictions and numerical simulations by almost one
order of magnitude. This discrepancy appears puzzling and a number
of questions remain unanswered. Their resolution requires further
investigations, which is the object of this study.

Since ionic liquids represent the asymptotic case of concentrated
electrolytes at zero dilution, we focus our study on ILs at this
extreme limit where electrostatic effects are most pronounced. A
rigorous experimental framework deciphers quantitatively how
measurements may deviate from theoretical predictions and numerical
simulations. Thus, our robust methodologies can guide further
experiments across the broader spectrum of concentrated electrolytes.

We measured surface forces using two complementary advanced SFAs
and two different ILs. This allowed the force-distance profiles to
be compared under varied quasi-static and dynamic conditions,
different confinement geometries (crossed-cylinders, sphere-plane),
different materials (mica, borosilicate) with molecularly smooth
surfaces of curvature spanning one order of magnitude (cm to mm).
We show that consistency can be established provided two criteria
are fulfilled. The hydrodynamic contribution to the net force is
admittedly crucial to be reduced but of much importance,
thermodynamic equilibrium must be approached during measurements.

\section*{Results and Discussion}

One primary mission of our investigation is to get force-distance
profiles under conditions as close as possible to thermodynamic
equilibrium all the way through from large distances down to
contact. The classical quasi-static SFA procedure translates the
lower mica surface of the apparatus by a small amount and then
allows it to come to rest (usually for no more than 10~s due to
intrinsic surface drifts) at the new position, which is expected to
have reached equilibrium~\cite{israelachvili1978}. The surface
separation, $D$, is then recorded by interferometry and the force,
$F$, between the two mica surfaces is calculated in the usual way,
from the difference in the change in displacement of each end of the
cantilever. The procedure is repeated to generate the entire
force-distance profile. However, any mechanical and thermal drifts
prevent genuine equilibrium position to be set at each successive
step, especially when the liquid medium is viscous as ILs are.
These limitations encountered in all previous SFA reports for ILs
are overcome by the high stability of our advanced homemade SFA
where precise nano-controlled displacement and fine variation of $D$
allow the effective movement to be lowered down to 9~pm/s and
equilibration times to be extended up to 90~s at any surface
separation over a few micrometers range from contact. These
experimental inputs offer a great platform for measuring IL surface
forces. Thus, scan durations can last as long as $\sim$10~h for
just a single one-way approach starting at $D \sim 500$~nm. Over
the same scan the surface motion can be actuated in a variety of
traveling speeds, dwell times (surfaces left to rest over a
predefined duration before the next nanosized step is initiated),
and incremental steps in the range 2 to 0.2~nm (all details in
SI~Appendix, section~4). With this procedure, hereafter referred to
``stepwise protocol'', both the surface separation and the force
relax together toward their equilibrium values after each
displacement step of the piezo-driven confining surface. As a
consequence, the hydrodynamic contribution can be substantially
reduced compared to continuous velocity methods~\cite{chan1985} at
equivalent speeds. Moreover, the challenging task of determining a
no-slip plane position becomes a nonissue in the force-distance
profile analysis. Thus, hydrodynamic, surface forces and structural
contributions can be unraveled using this appealing procedure. On
one hand, slowing down the movement reduces the hydrodynamic
contribution to the force since it grows as $1/D$ upon confinement
within a macroscopic continuum Reynolds framework for film
drainage~\cite{chan1985}. Concomitantly, with surfaces better
stabilized at each new gap by lengthening the equilibration time,
one favors the confined IL under stress to undergo its structural
rearrangements.

\begin{figure*}[!ht]
  \centering
  \includegraphics[width=\linewidth]{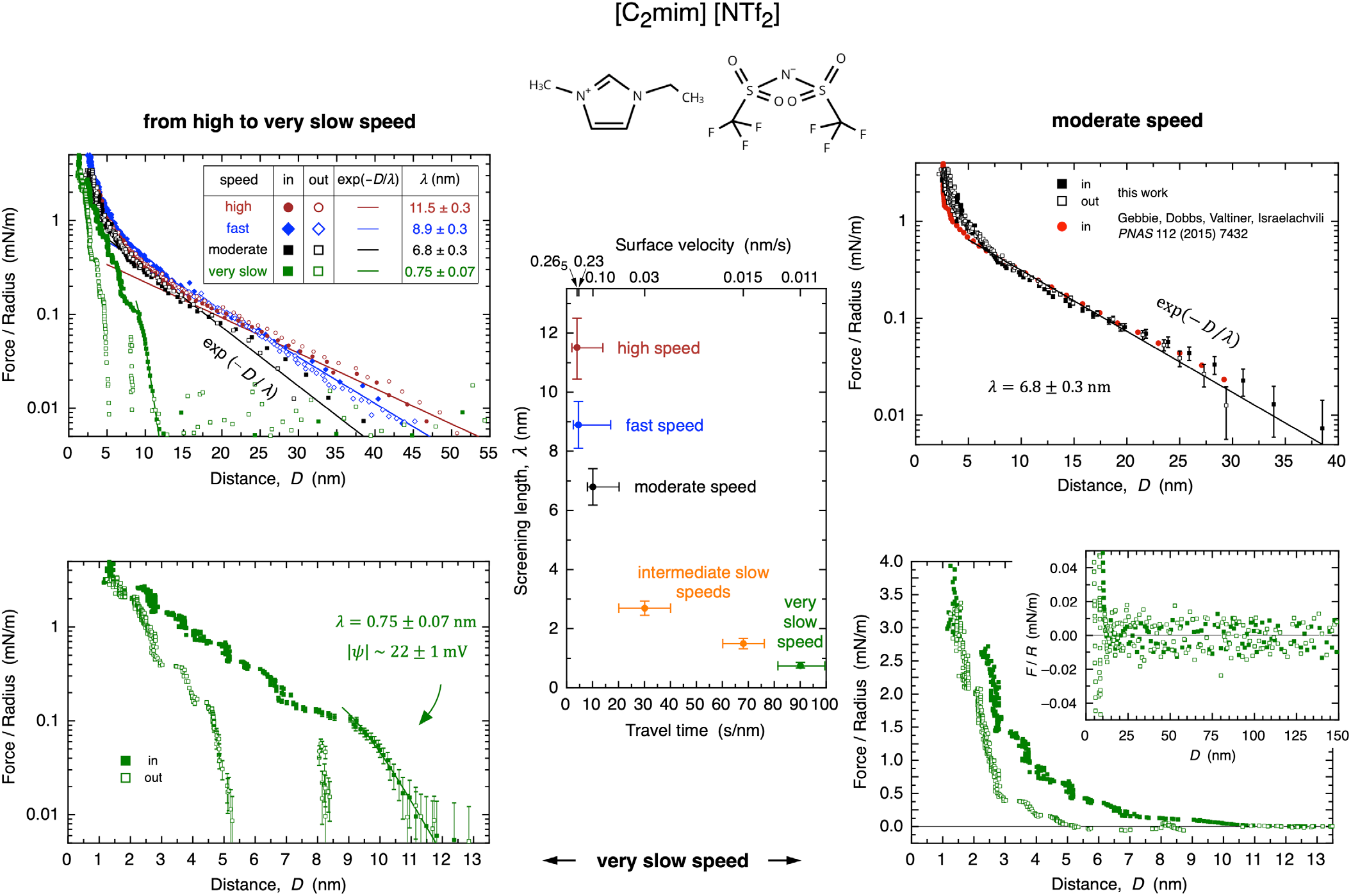}
  \caption{%
    Measured forces for 1-ethyl-3-methylimidazolium
    [C$_2$mim]$^+$ bis-(trifluoromethylsulfonyl)imide
    [NTf$_2$]$^-$, normalized by the mean radius of curvature of
    the surfaces between crossed mica cylinders, as a function of
    surface separation, $D$. Solid symbols correspond to the force
    on approaching the surfaces, while the open symbols are the data
    measured upon separation. The force-distance profiles are
    composed of two distinct regimes with spatial extent and
    magnitude both depending on velocity, stepwise and resting time
    conditions for surface movements (all parameters reported in
    SI~Appendix, section~4). At short separations the oscillatory
    functions of distance are prominently related to the
    structuration of the IL that may not be reversible upon
    decompression. At large separations the interactions are all
    monotonic repulsive and reversible. Fit tests by exponentials
    (solid lines) give large decay-lengths $\lambda$ in agreement
    with Gebbie et al.~\cite{gebbie2015} for the same IL in similar
    conditions (Top Right). However, increased slowdown leads to a
    progressive fading of the apparent long-range interaction with
    its related $\lambda$ being reduced by one-order of magnitude
    (central graph). The $\lambda$-uncertainties represent the
    estimated SD inferred from the fits on different force profiles
    collected during several approach and separation cycles and from
    possible variations in the repeatability of multiple independent
    experiments. The travel time segments indicate the $\tau$-ranges
    of displacement increments used along the measured monotonic
    repulsion only. At very slow speed (Bottom) the interaction is
    of short-range with $\lambda = 0.75 \pm 0.05$~nm close to the
    expected IL screening length suggesting an electrical
    double-layer interaction at constant potential $22 \pm 1$~mV
    boundary condition (solid line: DLVO fit).
  }
  \label{fig:1}
\end{figure*}

Figure~\ref{fig:1} presents a series of force-distance profiles
measured for the ionic liquid [C$_2$mim][NTf$_2$] (viscosity
$\eta = 37$~mPa$\cdot$s and IL characteristics: see SI~Appendix,
Table~S1) using our stepwise protocol at different velocities and
resting times (all experimental parameters are detailed in
SI~Appendix, section~4). The qualitative designations assigned for
high, fast, moderate, and very slow speeds correspond to a global
run time of about 20~min, 40~min, 1.5~h, and 10~h respectively for
a profile measured on approach of the two surfaces (solid symbols)
over a $\sim$500~nm range down to their contact. These total scan
durations are of course not proportional to an effective surface
velocity over the entire $D$-range since in the oscillatory regime
encountered at short separations a lot of time is taken up
traveling very little or no distance. Conversely at large
separations where strong short-range forces are not present, our
procedure allows an average speed to be defined (values reported in
the central graph). Figure~\ref{fig:1} compares the force-distance
profiles when at large separations this average speed
($\sim$0.27~nm/s at ``high speed''; brown circles) is reduced
systematically down to 0.011~nm/s at ``very slow speed'' (green
squares). In each case a repulsive force is observed which increases
in overall magnitude as the surface separation is reduced. The
monotonic continuous dependence with $D$ is broken with further
confinement signaling the occurrence of an oscillatory regime.
Below $D \sim 10$~nm, the force-distance profiles appear
discontinuous because of the intrinsic mechanical instability of
the cantilever that supports one of the mica surfaces: when the
gradient of the force is larger than the spring stiffness $K$
(i.e., $\partial F/\partial D > K$) the system jumps from unstable
to stable regimes, leaving unexplored and inaccessible
regions~\cite{israelachvili1978}. Some of those could be recovered
by measuring both the interaction upon approach (solid symbols) and
separation (empty symbols) of the two surfaces, but caution is
advised as the force curves differed in several ways as set out
hereafter.

Both the spatial extent and form of the oscillatory regime depend
on motion conditions of the mica surface that confines the IL. The
slower velocity the larger extent: the first oscillations could be
observed at $D$ as large as $\sim$8~nm for very slow approaching
surfaces but at gaps less than 5~nm at high speeds. The repulsive
background on which the oscillations superpose results from the
interplay of two effects. Faster movements induce larger
hydrodynamic repulsions, but instead give the confined IL less time
to undergo structural rearrangements and layering buildup. As a
result, the compressive energy barrier to squeeze out any layer in
formation is lowered and the related oscillation in the force
profile reduces its height or may even disappear (more details in
SI~Appendix, section~5). Such scenarios already observed in other
fragile systems such as sponge and lamellar
mesophases~\cite{freyssingeas1999,herrmann2014} once again underline
the importance of conducting force measurements under conditions as
close as possible to thermodynamic equilibrium.

Another feature is nonreversibility in the oscillatory regime upon
reversal of the movement, even at the slowest possible velocity as
illustrated by the semilog and linear scales of the force-distance
profile that was continuously measured upon approach of the surfaces
and then upon separation (Fig.~\ref{fig:1}, bottom). Contrary to
outward jumps with no regular positions, the location of the inward
jumps as a function of their rank becomes almost linear upon
increasing slowdown: the periodicity, $d$, approaches a value
around 0.8~nm as inferred from the slope of the line of best fit.
This $d$ value is in agreement with the short-range ordering
periodicity reported previously by AFM~\cite{wang2016} and
SFA~\cite{smith2015} and corresponds to ion pair dimensions. Several
arguments (SI~Appendix, section~5) harmonize with respect to
claiming that the IL film is microstructured with bare mica, becomes
anisotropic with layering buildup upon confinement, but does not
behave as a smectic phase. Plastic contributions, dependence in
confining wall velocity, different activation barriers, delays, and
transit times to squeeze out or incorporate a layer in confined
medium involving structural rearrangements and energetic changes,
all build a set of consistent observations that conspire for
caution in the usual assumption to considering surface forces and
hydrodynamic contribution as additive interactions for interpreting
total measured forces.

Contrary to the hysteresis observed in the oscillatory regime the
forces are reversible at separations where solvation forces are
absent. Reversing the surface motion at any stage gives the same
monotonic repulsion with characteristics not changing much upon
approach or separation. Interestingly, the force-distance profile
previously reported by Gebbie et al.~\cite{gebbie2015} for the same
[C$_2$mim][NTf$_2$] was reproduced when similar approach rates were
used (Fig.~\ref{fig:1}, top right with $\sim$1.5~h for a
$\sim$500~nm range and average speed $\sim$0.1~nm/s at large
separations). Indeed, these authors actuated the surface
displacement at constant rates of 0.2 to 1~nm/s implying that
their compressive force run lasted no longer than 1~h for bringing
the surfaces from $\sim$200~nm down to mechanical contact. It is
informative to follow their data analysis by attempting the large
separation regime to be described as an exponential repulsion.
Within this framework a large decay-length $\lambda = 6.8 \pm
0.3$~nm is inferred from the line of best fit to the data despite
scatter and force spanning over one order magnitude only. Similar
attempts of exponential fits for each velocity provide various
decay-lengths, all of large values, but with a strong dependence on
surface motion conditions.

Several arguments combine to not assign what would appear to be a
long-range exponential repulsion to an electrostatic double-layer
interaction, contrary to what was previously claimed for force
profiles measured in
ILs~\cite{espinosa2014,gebbie2015,smith2015,smith2016,lee2017,gebbie2017,lhermerout2018}.
First, fits to large scattered force data and besides over one
order magnitude only may cast doubt on the validity of an
exponential profile representation. Second, the force dependence on
surface motion conditions over a regime where solvation forces are
absent is at odds with a pure electrical double layer interaction.
Indeed, lower velocities with concomitant longer waiting times at
each new surface position reduce both magnitude and range of the
repulsions. The profiles decay back to zero force at increasingly
shorter distances with effective $\lambda$-values decreasing by more
than one order of magnitude from $\sim$11.5 to $\sim$0.75~nm at the
slowest possible surface displacement. The spatial extent becomes
quite narrow with no measurable force detected beyond 12~nm within
$\pm 0.01$~mN/m resolution (Inset) whereas the 8 to 9~nm signals
the transition to the solvation regime. With a force magnitude
reduced by more than one decade, the repulsion consequently appears
no longer long-ranged. The decay-length $0.75 \pm 0.05$~nm, now
close to the Debye-screening length expected for IL systems,
suggests that the asymptotic repulsion arises from an electrical
double layer interaction. Using the Poisson--Boltzmann (PB) equation
a fit would suggest an interaction at constant potential
$|\psi| \sim 22 \pm 1$~mV boundary condition at the onset of the
oscillatory regime transition. Nevertheless, caution must be taken.
The presence of the ionic layers makes difficult a genuine
determination of the double-layer origin. Furthermore, in more
accurate theories the screening of the charge near each surface is
more effective than in the PB mean field approach. As a result,
DLVO fits produce effective values smaller than the actual ones,
with the only exception to this behavior occurring for electrolytes
with very large counterions~\cite{kjellander1986}. However, these
complications in no way affect the screening length of the
double-layer interaction at large separations.

To support and extend this result, additional experiments were
carried out with another ionic liquid [C$_4$mim][PF$_6$], using
another surface force apparatus (dSFA, see SI~Appendix), with a
different geometry (sphere-plane instead of crossed-cylinders), a
radius of curvature $R$ an order of magnitude smaller and different
surface materials (borosilicate instead of mica). The surface
displacement system is also different, since it relies on a
controlled expansion rate of the piezoelectric actuator rather than
on a step movement~\cite{garcia2016,barraud2017}. Only the
repulsive regime observed at large separations is reported here as
the short-range elastic and solvation parts of the interaction were
discussed previously~\cite{garcia2018}. Here again, the effect on
magnitude and range of the repulsion was investigated when the
approach velocity of the surfaces was decreased over two orders of
magnitude from 3~nm/s to 15~pm/s.

\begin{figure*}[!ht]
  \centering
  \includegraphics[width=\linewidth]{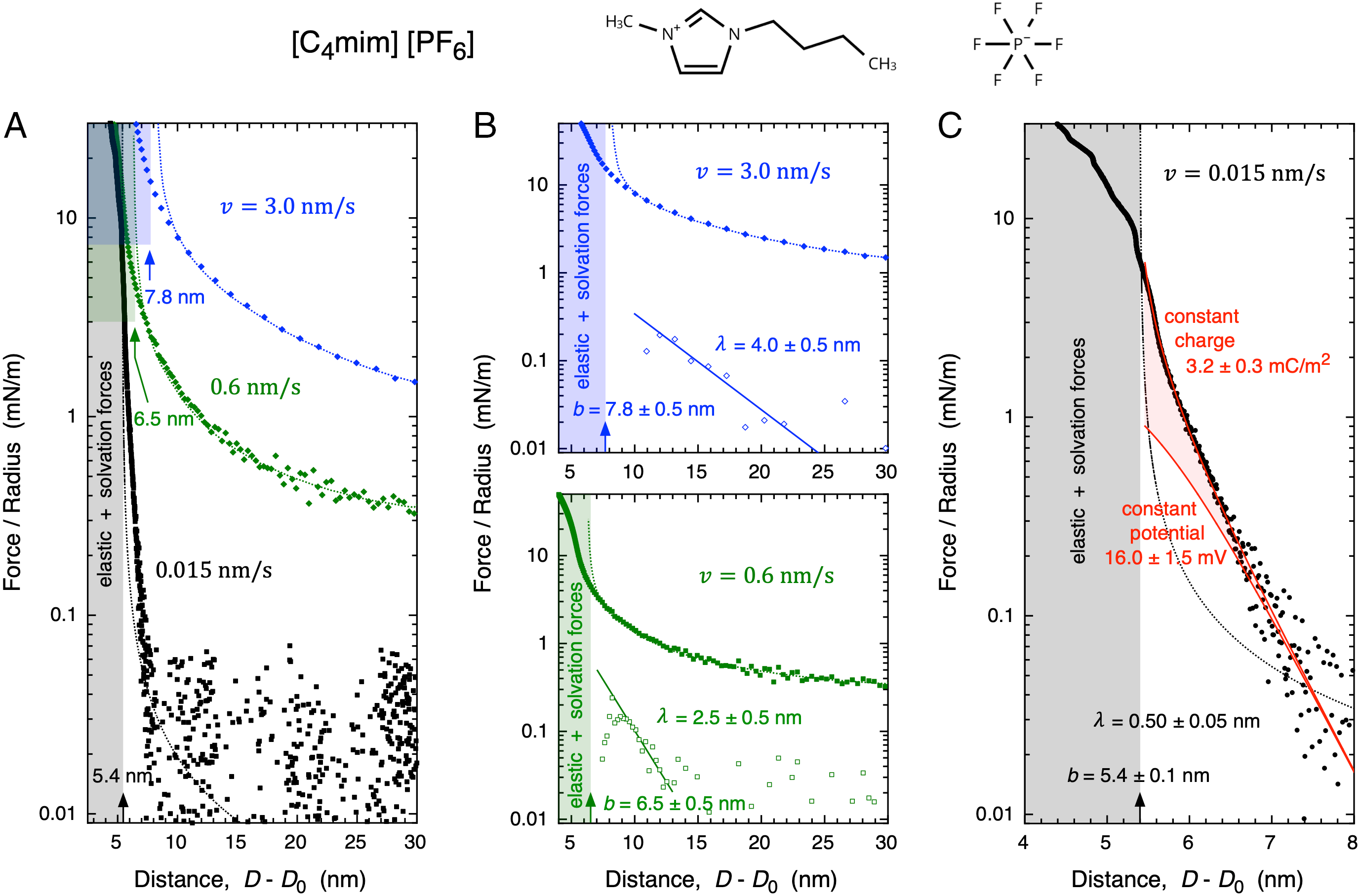}
  \caption{%
    Force-distance profiles of butyl-1-methyl-3-imidazolium
    [C$_4$mim]$^+$ hexafluorophosphate [PF$_6$]$^-$ between
    borosilicate surfaces in a flat-sphere (radius $R = 3.3$~mm)
    geometry.
    (A) Solid symbols are the measured forces at different
    approaching surface velocities $v = 3.0$ (blue), 0.6 (green),
    and 0.015~nm/s (black) while dotted lines represent the
    calculated corresponding hydrodynamic force (SI~Appendix,
    Figs.~S2 and S3).
    (B) Beyond the elastic and solvation forces regime (gray zone),
    the nonhydrodynamic interactions (open symbols), inferred by
    subtracting the hydrodynamics from the total measured force at
    $v = 3.0$ and 0.6~nm/s, could appear as having exponential
    profiles with decay-lengths $\lambda = 4.0 \pm 0.5$~nm and
    $2.5 \pm 0.5$~nm respectively.
    (C) At the slowest velocity $v = 0.015$~nm/s, the total
    interaction is short-range, exponentially decaying with
    $\lambda = 0.50 \pm 0.05$~nm close to the expected screening
    length for the IL, with a double-layer interaction at constant
    charge boundary condition.
  }
  \label{fig:2}
\end{figure*}

Figure~\ref{fig:2} reveals that at the highest velocity of
3.0~nm/s, the force profile (solid blue losanges) is dominated by a
hydrodynamic Reynolds behavior. The hydrodynamic force (dotted blue
line) is calculated with no fitting parameter using
$\eta = 157$~mPa$\cdot$s for the IL viscosity (SI~Appendix,
Table~S1) and a no-slip plane position at $b = 7.8 \pm 0.5$~nm
(SI~Appendix, Fig.~S2), an inferred value that agrees with
hydrodynamic measurements on similar systems~\cite{garcia2018}. Note
that only the relative displacements between the surfaces are
measured, and the $D_0$ position corresponds to the hydrodynamic
origin of the distances (SI~Appendix, Fig.~S1). This $D_0$
determination produces only a translation of the $D$-axis and has
no impact on the analysis of the interactions and the inferred
exponential decay lengths reported below. For intermediate
velocities (0.6~nm/s in green), the force profile exhibits two
distinct regimes. As long as the surface separations are
sufficiently large, the measurements align perfectly with a
hydrodynamic fit (dotted green line with same $\eta$, but
$b = 6.5 \pm 0.5$~nm, Fig.~\ref{fig:2}). However, as $D$ is
reduced below 15~nm the force profile deviates from a pure
hydrodynamic behavior indicating the competition with an additional
repulsion, that is further analyzed below. At the lowest
approaching velocity of 0.015~nm/s, the repulsion (black circles)
is short-ranged (no measurable force detected beyond $\sim$8~nm
within the rms $F/R$ resolution limit equal to 0.03~mN/m) with a
magnitude much larger than hydrodynamics (dotted black line).

To analyze further the force profiles, hydrodynamics and any other
possible contributions are considered as purely additive
interactions. Thus, the nonhydrodynamic component can be inferred
by subtracting the hydrodynamic contribution (dotted line) from the
total measured forces (solid symbols) as reported by open symbols
in Fig.~\ref{fig:2}B. This additional interaction is tempted to be
described by an exponential profile despite scattered data that
furthermore span over one-order of force magnitude only. Thus,
fitted decay-lengths $\lambda = 4.0 \pm 0.5$~nm and $2.5 \pm
0.5$~nm would be inferred respectively at the highest 3.0~nm/s and
intermediate 0.6~nm/s velocity.

Conversely, at the smallest 0.015~nm/s velocity, the force profile
exhibits a pure exponential behavior over two orders of magnitude
(Fig.~\ref{fig:2}C) with no need of any hydrodynamic subtraction
being much weaker. Its decay-length $\lambda = 0.50 \pm 0.05$~nm
is once again close to the Debye-screening length expected for IL
systems, suggesting that the asymptotic exponential repulsion arises
from a double layer interaction. In this distance regime, the van
der Waals attraction is weak [$F/R_\mathrm{vdW} = -A/6D^2$
neglecting retardation and electrolyte screening effects on the
effective Hamaker constant $A \sim 0.8 \times 10^{-20}$~J between
borosilicate substrates across the IL~\cite{shubin1993,mahanty1976}].
The profile is then fitted using a numerical solution of the
nonlinearized PB equation for constant charge and constant
potential boundary conditions. With the same caution taken
beforehand for marking the electrical double layer onset at the
transition with the oscillatory regime, the double-layer repulsion
follows a constant surface charge interaction
$|\sigma| = 3.2 \pm 0.3$~mC/m$^2$, a value in good agreement with
theoretical predictions~\cite{bazant2011}, whereas the surface
potential $|\psi| = 16.0 \pm 1.5$~mV amounts to a comparable value
for the other IL investigated differently with the other SFA.

\begin{figure}[!ht]
  \centering
  \includegraphics[width=\linewidth]{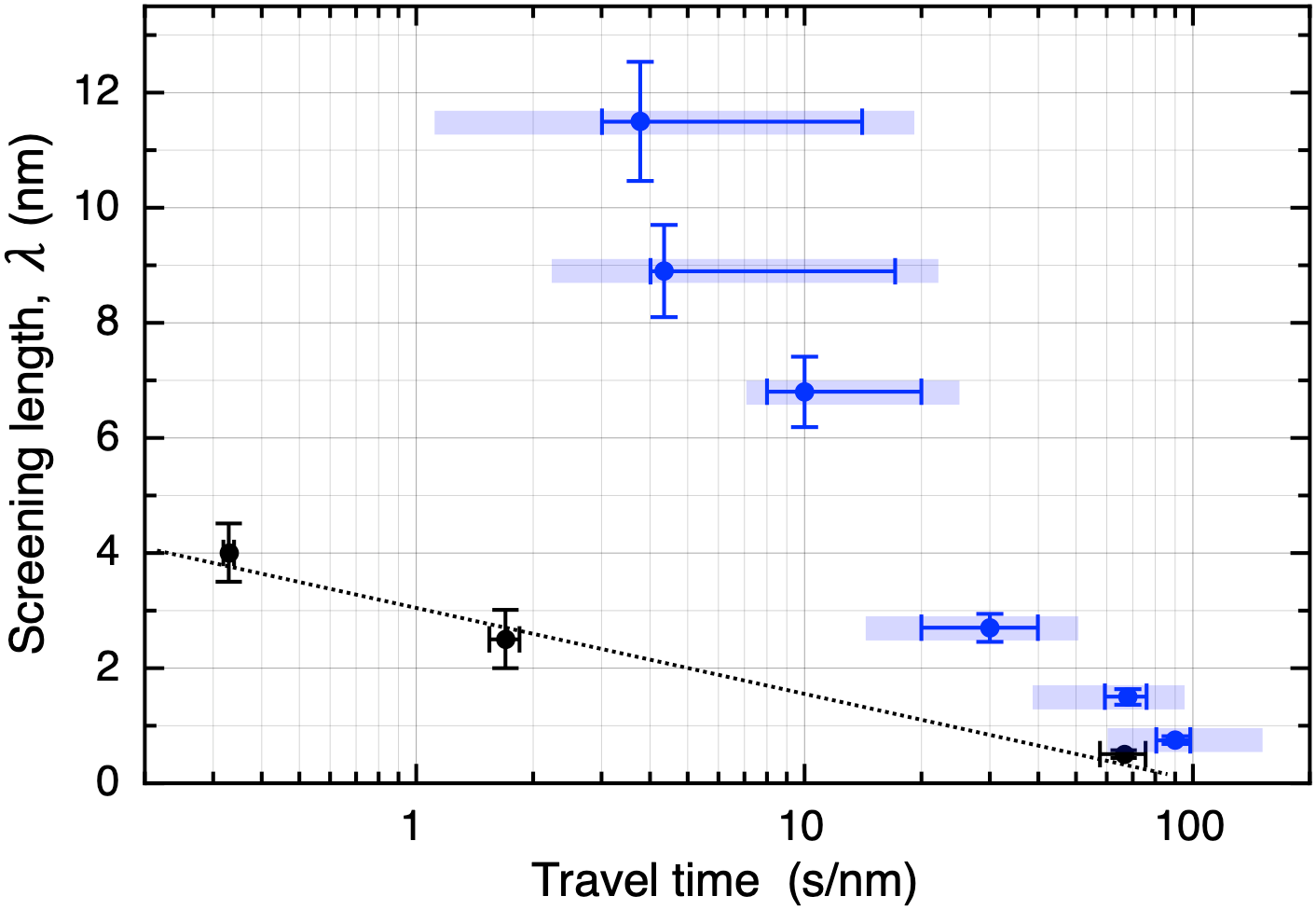}
  \caption{%
    Evolution of the apparent screening length, $\lambda$, of the
    monotonic repulsion at large separations (from
    Figs.~\ref{fig:1} and~\ref{fig:2}) as a function of travel time
    (inverse of approach velocity). Blue symbols represent
    [C$_2$mim][NTf$_2$] confined between crossed mica cylinders of
    radii of curvature $R \sim 10$ to 20~mm (all experimental
    parameters reported in SI~Appendix, Fig.~S4) using the stepwise
    protocol set for conventional SFA, while black symbols
    correspond to [C$_4$mim][PF$_6$] between borosilicate surfaces
    in a sphere-flat geometry ($R = 3.3$~mm) measured with the
    dSFA. Regardless of different ILs, surface materials,
    experimental configurations and protocols, both datasets show a
    steady decrease of $\lambda$ from $\sim$12~nm at high velocities
    (short travel times) down to 0.5 to 0.75~nm under quasi-static
    conditions (long travel times $\tau \sim 100$~s/nm). Error bars
    are uncertainties on $\lambda$ inferred from the fits and
    $\tau$-ranges of displacement increments used along the measured
    monotonic repulsion only (SI~Appendix, section~4), while
    shadowy blue rectangles indicate the full $\tau$-ranges used
    over the entire force-distance profile; the dotted line serves
    as a guide to the eye. Through two independent techniques and
    different ILs, the semilogarithmic representation reveals that
    the relaxation extends over two orders of magnitude in time
    scale, underscoring the IL slow progressive approach toward
    equilibrium. Thus, by allowing sufficient time for the confined
    systems to approach their equilibrium state, the decay lengths
    converge toward the theoretically predicted screening length.
  }
  \label{fig:3}
\end{figure}

Regardless of differences in ILs, surface materials, and
confinement geometries, both IL datasets demonstrate that the
screening length decreases steadily over two orders of magnitude in
travel time, dropping by more than one decade from $\sim$12~nm at
the fastest down to $\sim$0.5 to 0.75~nm at the slowest motion
conditions (Fig.~\ref{fig:3}). There, the measured $\lambda$ are
of the order of the ion sizes $a$ (SI~Appendix, Table~S1), with
$\lambda/a \approx 0.68$ to 1.25 for [C$_2$mim][NTf$_2$] and
$\approx 0.83$ to 1.08 for [C$_4$mim][PF$_6$] representing
short-range screening consistent with the theoretical predictions
by Attard~\cite{attard1993} and Jane\v{c}ek and Netz~\cite{janecek2009}.
The fact that $\lambda$ exceeds $\kappa_\mathrm{D}^{-1}$ by
$\sim$10 reflects the well-known finite-size corrections at high
concentrations but remains far from the anomalous
``underscreening'' regime previously reported by others (where
$\lambda/\kappa_\mathrm{D}^{-1}$ ratios up to $\sim$200 were
claimed). Both investigated ILs are in a regime where
$\lambda/a > \kappa_\mathrm{D}^{-1}/a$ remains well below the
Fisher--Widom threshold of $\sqrt{6} \approx 2.45$, whether one
considers the Debye length $\kappa_\mathrm{D}^{-1}$ or our
measured screening length $\lambda$. In such a regime, the
theoretical framework predicts an exponential (rather than
oscillatory) decay of the charge--charge correlations at large
separations together with no transition to complex screening
parameters~\cite{attard1993,janecek2009}. This is precisely what
we observe under quasi-static measurement conditions. The
oscillatory forces observed at short range ($D < 8$ to 10~nm)
reflect the structural layering of ILs at interfaces rather than
oscillatory electrostatic screening. These solvation forces arise
from the molecular structure of the liquid under
confinement---a distinct phenomenon from the charge--charge
correlations discussed in refs.~\cite{attard1993,janecek2009}. The
periodicity of these oscillations ($\sim$0.8~nm) corresponds to
ion-pair dimensions, confirming their structural origin. In
addition, the increase in surface layer thicknesses with slower
motions suggest that incomplete structural reorganization could
explain the observed deviations from equilibrium, possibly related
to capillary solidification~\cite{comtet2017}.

This study demonstrates that ionic liquid screening is short-range
under quasi-static conditions approaching thermodynamic equilibrium.
Such a conclusion would not have been possible if the experimental
framework for measuring force-distance profiles was not fulfilled
twofold. First, the use of two complementary techniques using most
advanced setups allows new procedures for accurate measurements.
Second, perturbations of the confined IL with transient structures
at the surface are also controlled and minimized once thermodynamic
equilibrium is reached. To fulfill this goal, a rigorous and
suitable protocol with two main experimental cautions was developed:
slow velocity for the moving surface, long equilibration times
between successive positions, all together with a subnano-controlled
way of the continuous varying thickness of the confined gap. This
twofold remark also leads to get insights into the entangled
contributions of hydrodynamic and surface forces to the total
measured interaction. Once the importance of measurements under
conditions approaching thermodynamic equilibrium is recognized, the
previous confused puzzle of wildly varying observations can be
rationalized. Our direct force measurements reconcile at last the
predictions set by theory and numerical simulations. The proposed
methodology sets the bases of a solid ground to pursue for
collecting valuable information concerning fine details of the
electrolytes behavior at elevated concentrations upon confinement.

Our investigation confirms the existence of a long-range force that
could approach a decaying exponential with distance. Nevertheless,
its related decay length, $\lambda$, is a strong function of travel
time (the inverse of the velocity) as revealed in
Fig.~\ref{fig:3}. The exceptional broad temporal range highlights
that ILs approach equilibrium upon confinement through slow,
progressive transitions rather than sharp ones. Such extended
relaxation dynamics spanning multiple time decades are reminiscent
of aging-like phenomena observed in a variety of systems, including
polymer films~\cite{priestley2005,charrault2008,napolitano2011,baumchen2012,mckenna2020},
historical paintings~\cite{bratasz2020}, granular
media~\cite{bocquet1998}. Thus, structural relaxation can extend
over years~\cite{priestley2005,charrault2008,napolitano2011,baumchen2012,mckenna2020}
or centuries~\cite{bratasz2020} at ambient temperature with
equilibration times strongly dependent on film thickness and
interfacial interactions~\cite{priestley2005,charrault2008,napolitano2011,baumchen2012,bocquet1998}.
In nanoscale polymer films, slow dynamics are associated with
hierarchical energy landscapes that offer a variety of pathways for
spatial and time relaxation through dynamical heterogeneities. As a
consequence, measurements were shown to depend critically on
equilibration protocols: insufficient waiting times lead to
artifacts and contradictory observations~\cite{baumchen2012}. While
the microscopic origins of the slow dynamics in confined ILs may
differ from those in polymer glasses, the phenomenological
similarity---extended relaxation over logarithmic time scales and
strong sensitivity to measurement protocols---suggests that future
investigations could benefit from concepts and theoretical
frameworks developed for these other interfacial systems.

The finding that screening dynamics in confined ionic liquids span
multiple time decades fundamentally challenges equilibrium-based
descriptions of these systems. Unraveling the physical origin of
this temporal complexity---whether structural frustration, precursors
to phase transitions, or novel mechanisms---represents a critical
frontier for understanding charge transport in nanoconfined
electrolytes and optimizing next-generation energy storage devices.

\section*{Materials and Methods}

Force-distance profiles were measured between pairs of atomically
smooth mica and borosilicate surfaces confining two aprotic ionic
liquids based on methylimidazolium cations:
1-ethyl-3-methylimidazolium bis-(trifluoromethylsulfonyl)imide
[C$_2$mim][NTf$_2$] and butyl-1-methyl-3-imidazolium
hexafluorophosphate [C$_4$mim][PF$_6$] at 25~$^\circ$C. Two
complementary homemade advanced surface force apparatus (SFA) were
employed. One is based on the conventional SFA
technique~\cite{israelachvili1978} operating with backsilvered mica
surfaces in a crossed-cylinder geometry using multiple-beam
interferometry (FECO) for distance measurement, $D$. The high
stability of our setup allows the moving surface to be stepwise
translated by a nano-controlled displacement and then left to rest
over a predefined duration before the next nanosized step is
actuated. Quasi-static equilibrium conditions can thus be approached
with effective slow movements as low as 9~pm/s and equilibration
times extended up to 90~s at any surface separation over a few
micrometers range from contact. The second complementary setup
(dSFA) provides force-distance profiles like conventional SFA but
also characterizes the linear harmonic force response of the thin
films confined between borosilicate surfaces in a sphere-flat
configuration. Both the relative surface displacement and the
deflection of a flexure hinge are simultaneously monitored by
Nomarski interferometry under continuous
approach~\cite{garcia2016,barraud2017}. With surface velocities that
can be varied over two orders of magnitude (from 3 down to
0.015~nm/s) rate-dependent effects can be distinguished from
equilibrium interactions. Detailed experimental procedures are
provided in the Supplementary Information. These include materials
characterization, distance calibration protocols, hydrodynamic
analysis, and how the different force contributions to the total
force can be disentangled.

\paragraph*{Data availability.}
All study data are included in the article and/or SI~Appendix.
Raw experimental data files have been deposited in
Zenodo~\cite{cross2026zenodo}.

\section*{Acknowledgments}

Romain Lhermerout is gratefully acknowledged for discussions on
the hydrodynamic aspects of ionic liquids under confinement. The
project was supported by the French National Research Agency (ANR)
through Grant No.\ ANR-19-CE30-0012. Technical development of the
advanced quasi-static Surface Force Apparatus beneficiated from the
experimental work carried out within the framework of the MICASOL
platform set between Institut Charles Sadron, CNRS, Strasbourg,
France, and SOLEIL synchrotron radiation facility, France
(\url{https://www.ics-cnrs.unistra.fr/micasol}).


\bibliography{references}
\bibliographystyle{apsrev4-2}

\end{document}


\title{Supporting Information for:\texorpdfstring{\\}{}
\textit{Short-range electrostatic screening in ionic liquids\texorpdfstring{\\}{}
as inferred by direct force measurements}}

\author{Benjamin Cross}
\email{benjamin.cross@univ-grenoble-alpes.fr}
\affiliation{Universit\'e Grenoble-Alpes, CNRS,
Laboratoire Interdisciplinaire de Physique,
Grenoble Cedex 9 38041, France}

\author{L\'eo Garcia}
\affiliation{Universit\'e Grenoble-Alpes, CNRS,
Laboratoire Interdisciplinaire de Physique,
Grenoble Cedex 9 38041, France}

\author{Elisabeth Charlaix}
\affiliation{Universit\'e Grenoble-Alpes, CNRS,
Laboratoire Interdisciplinaire de Physique,
Grenoble Cedex 9 38041, France}

\author{Patrick K\'ekicheff}
\email{patrick.kekicheff@ics-cnrs.unistra.fr}
\affiliation{Institut Charles Sadron, Universit\'e de Strasbourg,
CNRS UPR22, Strasbourg Cedex 2 67034, France}

\maketitle

\tableofcontents

\section{Materials}

Two common aprotic ionic liquids, both based on methylimidazolium
cations, but with different anions are compared. The cations differ
by the length of their hydrophobic alkyl chain,
1-ethyl-3-methylimidazolium, abbreviated as [C$_2$mim]$^+$, and
butyl-1-methyl-3-imidazolium noted [C$_4$mim]$^+$ (or [Bmim]$^+$),
whereas their associated anions are
bis-(trifluoromethylsulfonyl)imide ([NTf$_2$]$^-$) and
hexafluorophosphate [PF$_6$]$^-$ respectively. The
[C$_2$mim][NTf$_2$] and [C$_4$mim][PF$_6$] compounds were all
obtained from Solvionic (France), and used without further
purification. The [C$_2$mim][NTf$_2$] was dried under vacuum at
100~$^\circ$C for two days before deposition as a droplet
(50--100~$\upmu$L) in between the two confining extended cylinders
of the SFA. Dry nitrogen was flowed through the gas-tight injection
port of the setup at a pressure higher than the ambient atmospheric
pressure for a few hours before the apparatus was resealed. All
experiments were performed at 25~$^\circ$C in the presence of a
separate communicating reservoir located inside the sealed chamber
containing desiccant phosphorous pentoxide (P$_2$O$_5$). Each
experiment lasted between 24 and 72~h, and the P$_2$O$_5$
reservoirs were found to be unsaturated at the completion of each
experiment.

Table~\ref{tab:S1} summarizes the characteristics for these ionic
liquids, as collected from literature.

\begin{table*}[!ht]
\caption{Characteristics at 25~$^\circ$C for the two investigated
ionic liquids [C$_2$mim][NTf$_2$] and [C$_4$mim][PF$_6$].}
\label{tab:S1}
\begin{ruledtabular}
\begin{tabular}{llcc}
 & &
  [C$_2$mim]$^+$[NTf$_2$]$^-$ &
  [C$_4$mim]$^+$[PF$_6$]$^-$ \\
\hline
\multicolumn{2}{l}{Molecular weight $M$ (g/mol)} & 391.3 & 284.2 \\
\multicolumn{2}{l}{Viscosity $\eta$ (mPa$\cdot$s)} &
  37$^{[a]}$ & 157$^{[a]}$ \\
\multicolumn{2}{l}{Density $\rho$ (g/cm$^3$)} &
  1.52$^{[b]}$ & 1.37$^{[b]}$ \\
\multicolumn{2}{l}{Ionic density $\rho_\mathrm{ion}$ (mol/L)} &
  3.88$^{[c]}$ & 4.82$^{[c]}$ \\
\hline
\multirow{5}{*}{\rotatebox{90}{\shortstack{\small Structural\\ sizes (nm)}}} &
  Cation (nm) &
  $0.99\times0.50\times0.34^{[d]}$ &
  $1.20\times0.60\times0.35^{[d]}$ \\
 & Anion (nm) &
  $1.15\times0.48\times0.46^{[d]}$ &
  $0.59$ (diam.)$^{[e]}$ \\
 & Avg.\ ion pair size $a$ &
  $0.81^{[f]}$ & $0.75^{[f]}$ \\
 & Corr.\ dist.\ between like-ions &
  $0.90^{[g]}$ & $0.63$--$0.66^{[h]}$ \\
 & Nanoscale heterogeneities &
  expected$^{[i]}$ & $1.2^{[i]}$ \\
\hline
\multicolumn{2}{l}{Relative dielectric constant $\varepsilon_\mathrm{r}$} &
  12.3$^{[j]}$ & 11.4$^{[j]}$ \\
\multicolumn{2}{l}{Debye length $\kappa_\mathrm{D}^{-1}$ (nm)} &
  0.061$^{[k]}$ & 0.053$^{[k]}$ \\
\multicolumn{2}{l}{Refractive index} &
  1.422--1.423$^{[l]}$ & 1.417$^{[l]}$ \\
\end{tabular}
\end{ruledtabular}

\smallskip
\begin{minipage}{\linewidth}
\footnotesize
\flushleft$[a]$ Viscosity values for [C$_2$mim][NTf$_2$] \cite{yu2012,froba2008}
and for [C$_4$mim][PF$_6$] \cite{yu2012}.\\
$[b]$ Density at atmospheric pressure inferred from experiments
\cite{froba2008,montalban2015} and coarse molecular dynamics
simulations for [C$_4$mim][PF$_6$] \cite{bhargava2007}.\\
$[c]$ Ionic density calculated as $\rho_\mathrm{ion} = \rho/M$.\\
$[d]$ Dimensions estimated from the van der Waals radii of the
constituent atoms.\\
$[e]$ [PF$_6$]$^-$ anion size diameter from WAXS and numerical
simulations \cite{reichert2007}.\\
$[f]$ Average ion pair size $a$ calculated from the density of the
IL, assuming the packing fraction $\chi = (\pi/6)\rho a^3$ of 0.64.
This value corresponds to random packing of spheres \cite{song2008}
--- an assumption taken by Lee et al.\ \cite{lee2015} having noted
that the packing fraction is roughly constant for ionic liquids
\cite{slattery2007}.\\
$[g]$ Correlation distance between two cations or between two anions
inferred from WAXS and WANS for [C$_2$mim][NTf$_2$] \cite{fujii2008}.\\
$[h]$ Correlation distance between two cations in [C$_4$mim][PF$_6$]
of 0.63~nm inferred by WAXS \cite{billard2003} and of 0.66~nm
inferred from WANS \cite{triolo2006}.\\
$[i]$ Nanoscale structural heterogeneities arising from the
segregation of alkyl tails into oily domains that are embedded in a
charged, three-dimensional network \cite{canongia2006}. They were
evidenced in [C$_4$mim][PF$_6$] \cite{triolo2008} but not clearly
detected in [C$_2$mim][NTf$_2$] \cite{xiao2009}.\\
$[j]$ Relative dielectric constant for [C$_2$mim][NTf$_2$]
\cite{weingartner2008} and for [C$_4$mim][PF$_6$]
\cite{weingartner2008,mizoshiri2010}.\\
$[k]$ Debye screening length,
$\kappa_\mathrm{D}^{-1} = (\varepsilon_0\varepsilon_\mathrm{r}k_\mathrm{B}T /
2\rho_\mathrm{ion}e^2)^{1/2}$
calculated under the assumption that all ions are free and behave
as effective charge carriers (full dissociation); $k_\mathrm{B}$
designates the Boltzmann constant, $T$ the temperature, $\varepsilon_0$
the vacuum permittivity, $\varepsilon_\mathrm{r}$ the relative
permittivity (dielectric constant) and $e$ the elementary charge.\\
$[l]$ Refractive index at the sodium line ($\lambda_\mathrm{D} = 589.3$~nm)
for [C$_2$mim][NTf$_2$] \cite{froba2008,montalban2015} and for
[C$_4$mim][PF$_6$] \cite{montalban2015}.
\end{minipage}
\end{table*}

\section{Force measurement procedures}

Due to the atomic resolution in surface separation accessible, the
Surface Force Apparatus (SFA) technique~\cite{israelachvili1972,israelachvili1978}
allows a direct measurement of the interaction between two macroscopic
solid bodies or extended surfaces (of cm$^2$ area) confining a
liquid as a function of their separation. Force-distance profiles
determination reduces to the measurement of two quantities sets,
namely, the force of interaction, $F$, between the two bodies, and
the width of the gap, $D$, that separates them. Thus, the
measurements are carried out keeping the chemical potential of all
the chemical entities in the liquid constant. For the ionic liquids
investigated in this work, the force-distance profiles were measured
by means of two homemade advanced SFA versions with enhanced spatial
and time resolution and automation, as described in detail
elsewhere~\cite{garcia2016,barraud2017,kekicheff2001,kekicheff2019}.
One is based on the initial version of the Tabor-Israelachvili
SFA~\cite{israelachvili1972,israelachvili1978} using the
crossed cylinder geometry of two back-silvered atomically smooth
thin mica sheets (1--3~$\upmu$m) glued down on hemicylinders (radius
of curvature $R = 10$--20~mm) aligned at right
angles~\cite{israelachvili1972,israelachvili1978,kekicheff2001,kekicheff2019}.
The other SFA, on which one can measure the linear harmonic force
response of a thin film and, for this reason, is named dynamic
Surface Force Apparatus (dSFA)~\cite{garcia2016,barraud2017} uses
a sphere ($R = 3.3$~mm)--flat geometry with atomically smooth
borosilicate surfaces (roughness measured by atomic force microscopy
of $\sim$0.2~nm over a 100~$\upmu$m squared scan). In both devices,
one surface is rigidly coupled to the force sensor (lower
hemicylinder held at the end of a double-cantilever
spring~\cite{kekicheff2001,kekicheff2019}; glass sphere attached to
a flexure hinge~\cite{garcia2016,barraud2017}) whereas the opposite
surface (upper hemicylinder~\cite{kekicheff2001,kekicheff2019};
planar surface~\cite{garcia2016,barraud2017}) is mobile, allowing
the separation between the two surfaces to be controlled and finely
varied over a range from tens of micrometers down to molecular
contact by means of a piezo element superimposed on the coarse
positioning range of a motorized micropositioner.

In both SFAs the surface separation measurement is determined by
interferometry. In the conventional SFA, the optical thickness of
the medium between the mica surfaces is measured with multiple-beam
interferometry~\cite{tolansky1973,israelachvili1973,horn1991}. The
two partly silvered mica sheets, glued silvered side down onto the
surface of two glass cylinders, form an optical cavity into which
white light is directed~\cite{israelachvili1972,israelachvili1978}.
Multiple reflections are undergone between the two metallic surfaces
(silver 50~nm thick; reflectivity $>$98\% in the green region of
the visible spectrum). Only certain wavelengths of the transmitted
light have appreciable intensity: the spectrum is comprised of a
series of fringes of equal chromatic order (FECO)~\cite{tolansky1973}
whose wavelengths are related to the thickness and the refractive
index of each layer in the interferometer (mica sheets, confined
liquid)~\cite{kekicheff2001,israelachvili1973,horn1991,kekicheff1994}.
Standard matrix multiplication methods were used to determine the
coefficients of reflection and transmission for any combination of
layers widths of specified optical properties at varying wavelengths,
and consequently fringe positions could be
deduced~\cite{kekicheff2001,kekicheff2019,kekicheff1994,born1959}.
The outcome is that an absolute zero determination in separation can
be determined to about 0.1~nm by bringing the two bare atomically
smooth mica surfaces into contact. The comparison of this initial
situation and where an IL is sandwiched shows shifts in FECO
wavelengths, from which the optical thickness of the confined IL
medium can be determined. Both width and refractive index can thus
be inferred simultaneously~\cite{israelachvili1973,horn1991,kekicheff1994}.
The surface separation, $D$, is determined to within 0.2~nm whereas
the refractive IL index was found to be independent of $D$ at
$\pm 0.01$ with a value close to the bulk value reported in
literature (Table~\ref{tab:S1}, ref.\ 1).

In the dSFA, one order higher resolution (typically 20~pm) is
obtained as both the relative displacement of the mobile surface
and the deflection of the flexure hinge are independently measured
by means of Nomarski interferometers~\cite{barraud2017}. Note that
the absolute zero separation is not directly determined, but rather
a hydrodynamic origin (see hereafter, Sec.~3.1).
Anyway, the valuable independent determination of both surface
separation, $D$, and force, $F$, is of great value in the dSFA
contrary to the conventional SFA, in which the cantilever
deflection (and hence $F$) is calculated in the usual way, as the
difference between the expected and observed deflection of the
cantilever spring, i.e.\ from the deviation of the driving distance
calibration determined at large separations by multiple beam
interferometry where interaction surface forces are
negligible~\cite{israelachvili1972,israelachvili1978,kekicheff2001,kekicheff2019}.
Another advantage of the dSFA setup is its intrinsic rigidity
allowing large forces up to several hundreds of mN/m to be measured
as well as mechanical deformation of the surfaces to be
investigated~\cite{garcia2016,barraud2017}. With a flexure hinge
stiffness of $\sim$5000~N/m, the resolution in force is 0.1~$\upmu$N
giving a normalized force $F/R$ to be measured within 0.03~mN/m. In
the conventional SFA, the converse situation is encountered: the
cantilever spring stiffness chosen not too rigid (50--200~N/m)
allows the smallest normalized force to be detected to within
0.003~mN/m, while the maximum reliably measurable force will depend
on the mechanical compressibility of the entire system. Typically,
surface deformations occur for applied loads larger than
8--15~mN/m~\cite{attard1992,parker1992} and $F/R$ becomes
meaningless due to the deformation of the glue beneath the mica
sheet; for that reason, data are only reported for smaller loads.
Through the use of the Derjaguin
approximation~\cite{derjaguin1934}, the force normalized by the
local radius of curvature of the surfaces, $F/R$, may be related to
the free energy interaction per unit area between two parallel, flat
surfaces, $E$:
\begin{equation}
  \frac{F_\text{crossed-cylinders}}{2\pi R} =
  \frac{F_\text{sphere-flat}}{2\pi R} = E
\end{equation}
since for the separation range considered ($D \lesssim 1$~$\upmu$m)
the distance between the surfaces is equivalent to that between a
sphere and a flat plate (since $D/R \lesssim 3\times10^{-4}$
typically throughout an experiment)~\cite{derjaguin1934}.

In both devices, repulsive forces are seen as a continuous
deflection away from contact and are limited only by the onset of
deformation of the surfaces evoked here before. This deflection
technique is not suitable for the measurement of strongly attractive
surface interactions because of the mechanical instability which
occurs when the slope of the attraction exceeds the spring constant
of the force measuring device. As a consequence, the technique
allows forces to be measured only over the regions where the
gradient of the force $|\partial F/\partial D|$ is smaller than the
spring constant $K$. In these measurements only parts of the force
curves are directly accessible and the force vs.\ distance profile
appears discontinuous, with jumps from unstable (i.e.\
$|\partial(F/R)/\partial D| > K/R$) to stable mechanical regimes.
Use of stiffer cantilever springs would increase the range of
stability, but were not used as they were at the expense of
sensitivity in measuring the force.

\section{Analysis of dSFA measurements}

In dSFA measurements, the independent determination of both surface
separation and force allows a detailed analysis of the different
contributions to the total measured force. This section describes
the successive determination of the reference frames for distance
measurements (hydrodynamic origin, Sec.~3.1) and
for the no-slip plane positions (Sec.~3.2). The
conditions for reliably separating surface forces from hydrodynamic
contributions are addressed in Sec.~5.2.

\subsection{Hydrodynamic origin of distances}
\label{sec:hydro_origin}

A procedure that drives the two surfaces toward each other at
constant speed can partially overcome the aforementioned
restriction. This approach enables continuous recording of both
repulsive and attractive surface forces, as previously demonstrated
in many systems using the classical SFA technique~\cite{chan1985}
and in ionic liquids (ILs) using the dSFA ([C$_4$mim][PF$_6$] in
ref.~\cite{garcia2018}). At very large separations, the surfaces
approach each other with uniform velocity. However, at smaller
separations, the cantilever supporting one surface begins to deflect
as intermolecular forces emerge. The hydrodynamic force is
calculated using a macroscopic continuum Reynolds framework for film
drainage, using the bulk value of the viscosity $\eta = 157$~mPa$\cdot$s
for [C$_4$mim][PF$_6$] (Table~\ref{tab:S1}, ref.~1).

\begin{figure}[H]
  \centering
  \includegraphics[width=0.5\linewidth]{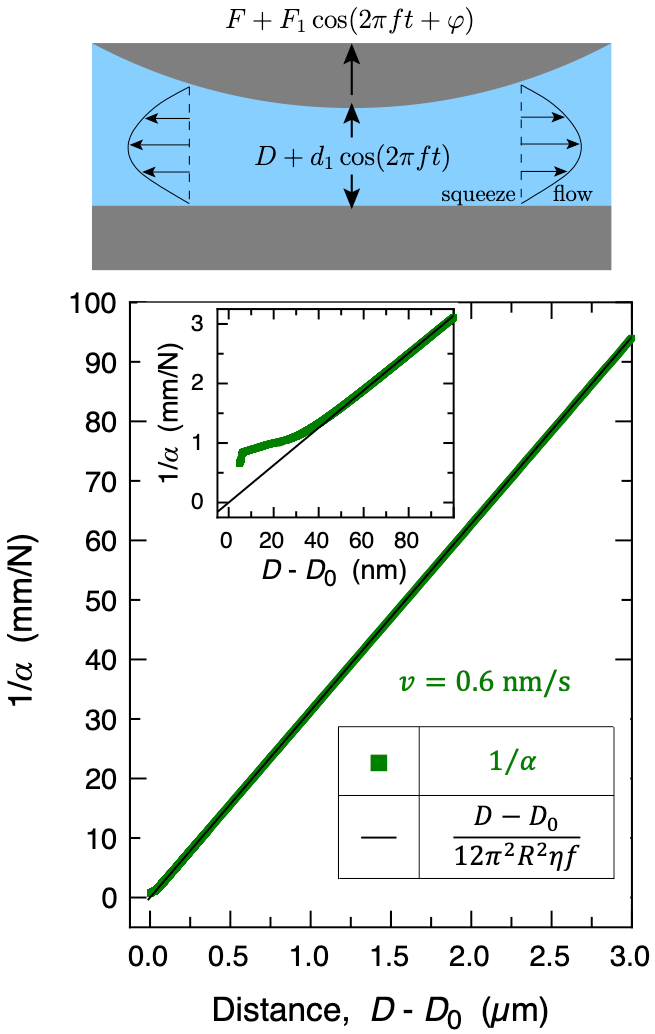}
  \caption{%
    Determination of the hydrodynamic origin $D_0$.
    \textit{Top:} Schematic illustration of a dSFA experiment with
    [C$_4$mim][PF$_6$] showing the various measured quantities.
    \textit{Bottom:} Inverse damping coefficient $1/\alpha$ as a
    function of surface separation (green squares). The hydrodynamic
    origin distance $D_0$ is determined by the intersection of the
    linear fit (black solid line, with $\eta = 157$~mPa$\cdot$s,
    $R = 3.3$~mm, and $f = 220$~Hz) with the distance axis.
    \textit{Inset:} Enlargement of the data at short separations
    showing the deviation from the linear behavior below 50~nm due
    to the elasto-hydrodynamic response caused by finite surface
    compliance.
  }
  \label{fig:S1}
\end{figure}

The high sensitivity of the dSFA allows a hydrodynamic origin of
the distances, $D_0$, to be set with a 1~nm resolution using a
sinusoidal flow. Thus, a sinusoidal displacement at frequency $f$
and amplitude $d_1$ is superimposed on the continuous linear
approach via the piezoelectric device (Fig.~\ref{fig:S1}). The
oscillations are maintained at values smaller than the nominal film
thickness $D - D_0$ (typically less than $0.01(D - D_0)$) to
ensure a linear response and to avoid perturbing the quasi-equilibrium
state (experimentally confirmed). The superimposed sinusoidal
displacement induces an oscillatory flow within the film, generating
a pressure field on the sphere that produces a sinusoidal force at
frequency $f$ with amplitude $F_1$ and phase shift $\varphi$
relative to the displacement.

The experimentally measured damping coefficient $\alpha$ corresponds
to the ratio $F_1\sin\varphi/d_1$. Within a linear regime with
no-slip boundary conditions at the solid-liquid interface, the
inverse viscous damping
$1/\alpha = (D - D_0)/(12\pi^2\eta R^2 f)$
increases linearly with the liquid film thickness~\cite{barraud2017}.
Figure~\ref{fig:S1} illustrates this relationship during continuous
reduction of the gap between the two borosilicate surfaces confining
the [C$_4$mim][PF$_6$] ionic liquid at a constant driving speed of
0.6~nm/s in the dSFA.

As evidenced in Figure~\ref{fig:S1}, there is a considerable
distance regime over which surface forces are negligible: the
force-distance profile conforms well to the theoretical predictions
for drainage of thin liquid films between extended solid surfaces.
At large separations, the linear behavior results from two
contributions only: the dominant hydrodynamic force and the
considerably much weaker van der Waals attraction. However, a
measurable deviation comes out when the gap thickness falls below
approximately 50~nm. This departure from linearity over this small
distance range results from an elasto-hydrodynamic response of the
confined viscous film, as it will be analyzed further below.

\subsection{No-slip plane position}
\label{sec:noslip}

\begin{figure*}[!hbtp]
  \centering
  \includegraphics[width=\linewidth]{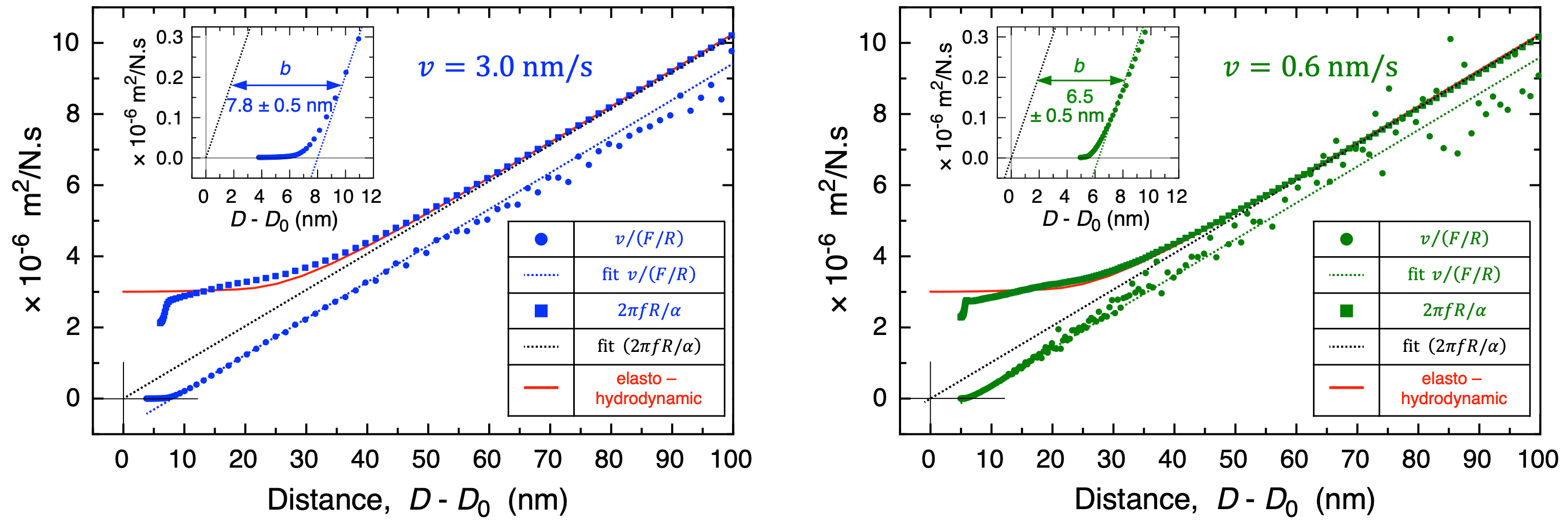}
  \caption{%
    Determination of the no-slip plane position for
    [C$_4$mim][PF$_6$] through hydrodynamic analysis at velocities
    $v = 3$~nm/s (left) and 0.6~nm/s (right). All in blue at
    $v = 3$~nm/s and in green at $v = 0.6$~nm/s, solid circles
    represent the experimental data $v/(F/R)$ for drainage at
    velocity $v$, squares are related to the harmonic quantities
    $2\pi f R/\alpha$ inferred from oscillatory measurements, while
    dotted lines indicate the theoretical predictions. The dotted
    black line is a linear fit to the $2\pi f R/\alpha$ data for
    film thicknesses beyond 100~nm, allowing $D_0$ to be determined
    (see also Fig.~\ref{fig:S1}). At every velocity, the identical
    slopes of both data sets confirm a constant viscosity
    independent of film thickness, while the difference in
    $D$-intercepts directly yields the no-slip plane position $b$
    (insets). The red solid line represents the elasto-hydrodynamic
    calculation for the confined IL between the elastic coated
    borosilicate surfaces.
  }
  \label{fig:S2}
\end{figure*}

At this stage, it is essential to examine how the elastic surface
layers (adsorbed IL) may significantly affect the interfacial
hydrodynamics. To this end, it is necessary to determine the
positions of the no-slip plane, denoted as $b$, for drainage flows
at all velocities used in the reported experiments. When a sphere
and a plane undergo a relative perpendicular motion at velocity $v$,
Stokes forces become significant. Thus, the ratio between the
velocity and the force (normalized by its radius of curvature),
namely $v/(F/R)$, will follow a linear relationship with the
distance as $(D - D_0 - b)/(6\pi\eta R)$, at least at large
separations where the hydrodynamic interactions dominate.
Figures~\ref{fig:S2} illustrate this expected behavior for the
velocities $v = 3$~nm/s (blue dots) and $v = 0.6$~nm/s (green
dots). The same Figures also report the experimental data
$2\pi f R/\alpha$ using the damping coefficient used to determine
$D_0$ (squares in blue for $v = 3$~nm/s and in green for $v =
0.6$~nm/s). These data align along a straight line (black dotted
line: linear fit carried out for film thicknesses larger than
100~nm) whose slope corresponds precisely to that of the drainage
quantity $v/(F/R)$. This slope, given by the ratio $1/(6\pi\eta R)$,
remains independent of the surface separation, demonstrating that
the liquid film has a constant viscosity regardless of its thickness
with the same value as in the bulk kept down to the shear plane.
Although the two data sets exhibit identical slopes, their linear
extrapolations intercept the separation axis at markedly different
points. The shift between these separation intercepts directly
corresponds to the no-slip plane onset $b$, allowing a precise
experimental determination of its position: $b = 7.8 \pm 0.5$~nm
and $6.5 \pm 0.5$~nm at respectively $v = 3$ and 0.6~nm/s
(Fig.~\ref{fig:S2} insets). At the lowest velocity $v = 0.015$~nm/s,
the hydrodynamic determination of $b$ is not feasible because the
hydrodynamic contribution is not dominant. Therefore, an alternative
approach was adopted: the position of $b$ was established by setting
it at the point where structural solvation forces begin, an onset
coinciding with the location of the adsorbed layers on the
borosilicate substrates.

The difference between $D_0$ determined from sinusoidal measurements
and $b$ obtained from continuous hydrodynamic analysis on the same
experiment indicates the presence of an IL layer that responds to
oscillatory flow while remaining immobile during steady drainage
flow~\cite{garcia2018}. While these layers remain immobile under
continuous drainage conditions, they actively participate in the
dynamic response of the system. The elasto-hydrodynamic
calculation~\cite{leroy2011} for a viscous film confined between
two elastic extended solid surfaces is reported as the solid red
line in Figure~\ref{fig:S2}. The numerical calculation assumes a
bulk viscosity value for the IL ($\eta = 157$~mPa$\cdot$s) and
Young's modulus of 64~GPa for each borosilicate surface and 18~MPa
for each coated layer 3.5~nm thick. Although this calculation
cannot capture all the details at short separations over the
ultimate 50~nm range, it successfully reproduces the main behavior
of the system under harmonic excitation. Thus, the hydrodynamic
plane position $b$ appears to correspond to twice the coated layer
thickness (each layer contributes to $b/2$), and the elastic modulus
value is consistent with previous direct surface force
measurements~\cite{garcia2018}.

Since the $D_0$ determination only translates the $D$-axis without
affecting the interaction analysis presented in the main text, all
force profiles are presented as a function of the relative surface
displacement.

The possibility that liquid viscosities are enhanced in very thin
films and deviate from their bulk values near surfaces has been
debated in the literature for decades. The current consensus appears
to rule out distance-dependent viscosity functions from solid
interfaces~\cite{bocquet2010}. The liquid is therefore assumed to
maintain a constant viscosity up to the shear plane, though this
plane need not coincide with the physical surface (as exemplified
in Fig.~\ref{fig:S2}). This plausible approach avoids invoking
microscopic explanations for enhanced viscosity near solid
surfaces~\cite{boumalham2010,ueno2010}, instead following the
simpler framework proposed in the pioneering work of Chan and
Horn~\cite{chan1985}: an immobile liquid region exists adjacent to
each solid surface during continuous drainage flow. As these authors
stated: ``this very simple-minded approach has two advantages. First,
it introduces only one adjustable parameter [$b$] with which to fit
the experimental results, and second, it leaves the viscosity $\eta$
independent of separation.''

\begin{figure*}[hbtp]
  \centering
  \includegraphics[width=\linewidth]{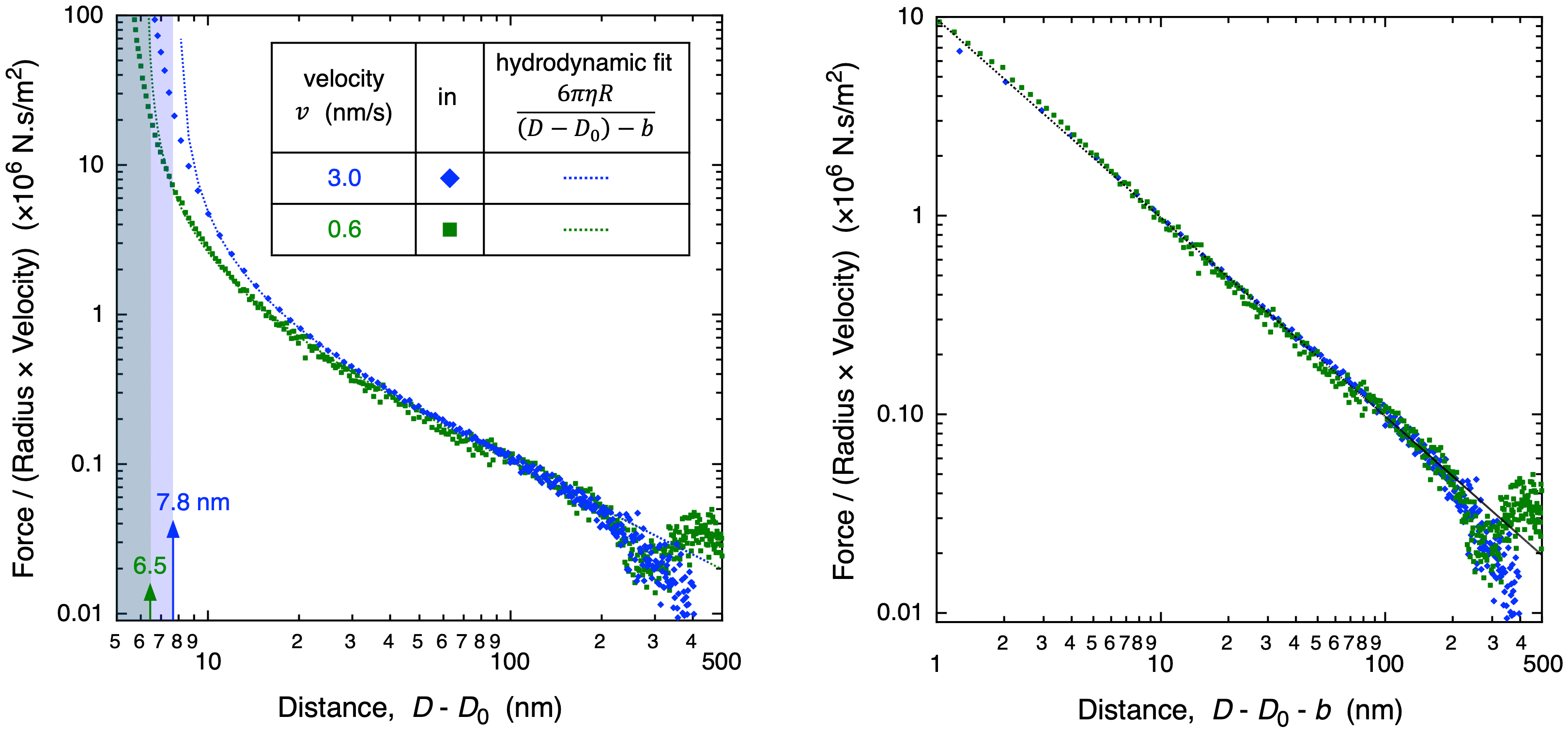}
  \caption{%
    Velocity-normalized force--distance profiles demonstrating
    near-universal scaling behavior.
    \textit{Left:} Force divided by radius and velocity, $(F/R)/v$,
    as a function of the liquid film thickness $(D - D_0)$ for two
    approaching velocities $v = 3.0$~nm/s (blue diamonds) and
    0.6~nm/s (green squares) in [C$_4$mim][PF$_6$]. Dotted lines
    represent hydrodynamic calculation using the theoretical
    expression $6\pi\eta R/(D - D_0 - b)$. The shadowy regions
    indicate the small separation range below 10~nm where the
    surface forces become significant. Vertical arrows mark the
    no-slip plane positions $b$ at $7.8 \pm 0.5$~nm (blue) and
    $6.5 \pm 0.5$~nm (green) as determined in Fig.~\ref{fig:S2}.
    \textit{Right:} after accounting for the no-slip plane offset,
    the data are well described by the theoretical hydrodynamic
    prediction over the entire measurable range, confirming a
    constant viscosity behavior that is independent of film
    thickness, at least down to the smallest separations where
    surface forces dominate. Although the data merge onto a single
    master curve over most of the distance range, slight differences
    still persist at small separations as revealed in Figure~2b
    (main text).
  }
  \label{fig:S3}
\end{figure*}

One can systematically compare results across different velocities
by normalizing the measured force $F$ by the velocity of the moving
surface (Figure~\ref{fig:S3}). At first approximation, all
experimental data appear to fall onto a single master curve,
regardless of whether the driving speed is maintained strictly
constant over the entire distance range or varied incrementally
through small step-size adjustments at specific gap separations.
The renormalized force profiles lie on a single curve: this suggests
a remarkable degree of similarity in the surface interactions across
the IL film, seemingly independent of the hydrodynamic contribution
magnitude. Such velocity-independent behavior has been previously
reported by other research groups investigating similar confined IL
systems (see for instance ref.~\cite{bocquet2010}).

However, the persistence of small discrepancies between velocity
conditions warrants cautious interpretation. These deviations,
though small, correspond to out-of-equilibrium effects that are
velocity-dependent (Figure~3, main text). This observation
necessitates a more nuanced approach to force decomposition, as
discussed in Sec.~5.1 for structural~forces.

\section{Analysis of conventional SFA measurements}

Conventional SFA measurements differ fundamentally from the
continuous velocity protocols employed in dSFA experiments described
in the preceding section. In usual SFA
measurements~\cite{israelachvili1972,israelachvili1978,kekicheff2001},
the force between the two surfaces in crossed-cylinder configuration
is inferred by expanding or contracting by a predefined
displacement, $\delta$, the piezoelectric crystal on which one
surface is attached, and then measuring interferometrically the
actual distance that the two surfaces have moved relative to one
another, $\Delta D$. Any difference in the two distances when
multiplied by the double-cantilever spring constant $K$ gives,
using Hooke's law, the difference between the forces at the initial
and the final separations, $\Delta F$:
\begin{equation}
  \Delta F = K(\delta - \Delta D)
  \label{eq:SFA_force}
\end{equation}

Using cantilevers with a spring constant, $K$, in the range
50--200~N/m gives an accuracy in force of 10~nN, corresponding to
a force normalized by radius, $F/R$, that could be as low as
0.002~mN/m for surfaces with radii of curvature, $R$, in the range
10--40~mm. However, this resolution has never been attained in any
SFAs, even in our advanced SFA version, despite enhanced spatial
and time resolution, automation, and thermal and mechanical
stability. Beyond intrinsic noise in the data measurements, the
main limit arises from the uncertainty of the zero of force. Its
determination is inferred from the driving distance calibration set
at large separations by multiple beam
interferometry~\cite{kekicheff2001,kekicheff2019,kekicheff1994}
where interaction surface forces are assumed to be negligible. As
a result, in this reported work, the lowest $F/R$ value, that can
be reliably measured, is about 0.005~mN/m.

As exemplified by Eq.~\eqref{eq:SFA_force}, the force
determination analysis faces specific challenges due to the stepped
displacement protocol. Because the two settings, namely, cantilever
displacement (i.e.\ force $F$) and surface separation $D$, cannot
be handled simultaneously and measured independently upon time
during the surface traveling, a genuine value of force is obtained
only and only if the cantilever position is at mechanical
equilibrium when the final separation is measured. Otherwise, the
measure is marred by a contribution of the unrelaxed confined fluid.

One way to overcome this drawback would be to collect SFA data in
the complementary situation in which a hydrodynamic force is
purposedly generated while being controlled and evaluated. Within
this framework proposed by Chan and Horn~\cite{chan1985}, one
surface is driven continuously at constant velocity and the surface
forces are inferred by subtracting a calculated hydrodynamic
contribution from the total measured force. In such experiments the
hydrodynamics and any other possible contributions are considered
as purely additive interactions. This method has been extremely
beneficial for a variety of observations embracing different liquids
and materials. Lhermerout and Perkin also applied it to ILs using
carefully designed surface trajectories $x(t)$ for modeling and
inferring the position of the no-slip plane, $b$~\cite{lhermerout2018}.
Their analysis demonstrated that hydrodynamic contributions were
significant, while still accounting for a residual long-range
interaction decaying exponentially with distance. However, in a
conventional SFA an accurate determination of the IL no-slip plane
position is prevented from implementing advanced protocols similar
to those set in dSFA (see preceding section). As a result, any
uncertainty in determining $b$ may induce significant errors when
inferring the non-hydrodynamic component through subtraction,
beyond the inherent questionable assumption of additivity. As it is
shown below, all these limitations no longer exist when the moving
surface is stepwise translated by a nano-controlled displacement
and then left to rest over a predefined duration before the next
nanosized step is actuated. With this procedure, hereafter referred
to ``stepwise protocol'', the relaxation of the confined liquid can
take place after each thickness change of the gap separating the
two surfaces. As a consequence, the hydrodynamic contribution can
be substantially reduced compared to continuous velocity methods at
equivalent speeds. Moreover, the challenging task of determining a
no-slip plane position, $b$, becomes a non-issue in the
force-distance profile analysis.

The stepwise SFA protocol remains of value provided the surface
separation change is quasi-static, that is the measurement is
carried out close to thermodynamic equilibrium. To our knowledge,
the quantitative assessment for which such a measurements method
is valid has never been established so far. Resolution of this
question requires further investigations which is the object of
this section.

Driven by the piezoelectric actuator, the discrete displacement
steps, $\delta$, lead to instantaneous force transients followed
by exponential relaxations as the confined liquid flows. As a
result, the opposite surface attached to the cantilever moves off
with a motion that relaxes upon time according to its stiffness,
$K$, and to the extent of the confining surfaces through their
radii of curvature, $R$. For a given liquid viscosity, $\eta$, the
motion response is nonlinear as it arises from the
distance-dependent damping coefficient
$\lambda(D) = 6\pi\eta R^2/(D - b)$, which is an inverse function
of the surface separation. This nonlinearity prevents the use of
standard analytical solutions and necessitates the development of
an iterative computational approach where the equation of motion is
solved individually for each displacement step.

\begin{figure}[hbtp]
  \centering
  \includegraphics[width=0.9\linewidth]{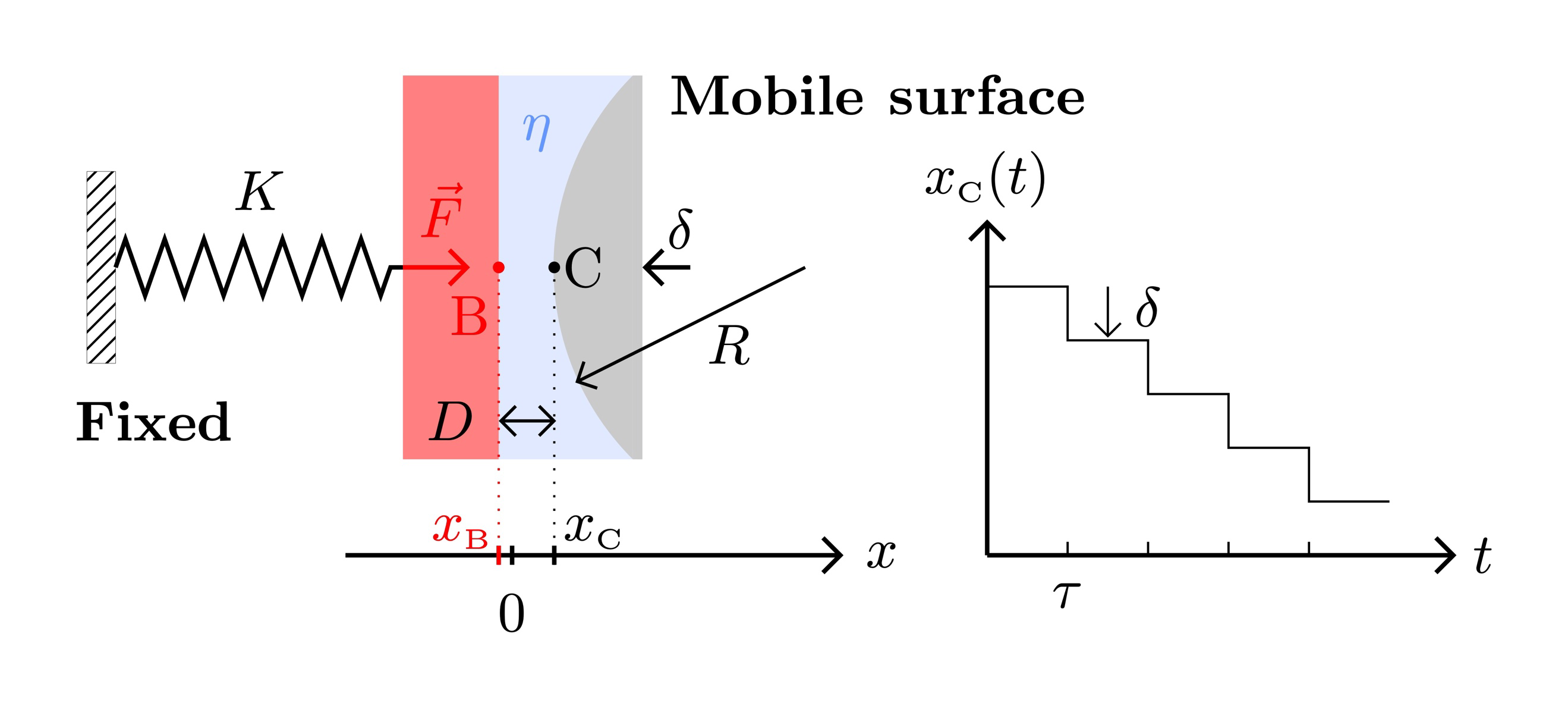}
  \caption{%
    Schematic representation of the nonlinear spring-damper system
    modeling the dynamic response of a conventional Surface Force
    Apparatus (SFA). The grey mobile surface carrying point C is
    controlled by the piezoelectric actuator, while the red surface
    carrying point B is connected to the cantilever attachment with
    spring constant $K$. The blue region represents the ionic liquid
    of viscosity $\eta$ confined between the two mica surfaces in
    crossed-cylinder configuration, equivalent to a flat against a
    sphere of radius of curvature $R$~\cite{derjaguin1934}. The
    stepwise displacement is composed of incremental translations
    $\delta$ of the mobile grey surface driven by the actuator at
    time intervals $\tau$. Each step increment $\delta$ creates an
    instantaneous response that is then followed by an exponential
    relaxation of the cantilever position via a distance-dependent
    damping coefficient $\lambda(D)$.
  }
  \label{fig:S4}
\end{figure}

To mimic the SFA experiments reported in the main text (Figure~1),
the numerical calculation considers a series of displacement steps
bringing the surfaces together. Rather than continuous motion, each
step is applied with an amplitude $\delta$ ($< 0$ for surface
approach, as schematized in Fig.~\ref{fig:S4}), at regular time
intervals $\tau$. The initial surface separation is set to $D(0)$
with point C at position $x_\mathrm{C}(0) = D(0)$ and the
cantilever at $x_\mathrm{B}(0) = 0$, representing no initial
deflection.

The motion of the surface carrying the point B is governed by a
force balance between the viscous Reynolds force
$F_\mathrm{Reynolds} = \lambda(D)\cdot\dot{D} =
\lambda(D)\cdot(\dot{x}_\mathrm{C}(t) - \dot{x}_\mathrm{B}(t))$
from the confined fluid and the elastic force
$F(t) = -K\cdot x_\mathrm{B}(t)$ from the cantilever. The
differential equation governing the motion of point B is then
written as:
\begin{equation}
  \lambda(D)\frac{dx_\mathrm{B}(t)}{dt} + K x_\mathrm{B}(t)
  = \lambda(D)\frac{dx_\mathrm{C}(t)}{dt}
  \label{eq:eom}
\end{equation}

The stepped displacement of point C follows
$x_\mathrm{C}(t) = D(0) + \sum_{i=1}^{N}\delta\cdot
H(t - (i-1)\tau)$, where $H$ denotes the Heaviside step function.
The velocity of C consists of a series of Dirac impulses:
$dx_\mathrm{C}(t)/dt =
\sum_{i=1}^N\delta\cdot\delta_\mathrm{Dirac}(t - (i-1)\tau)$.
This mathematical representation captures the essence of the
experimental protocol where the discrete positioning steps create
instantaneous velocity spikes followed by relaxation periods.

The surface motion response is composed of two distinct phases for
each time interval $[(i-1)\tau, i\tau]$. During the impulse phase
at each instant $t_{i-1}$, the Dirac impulse on C velocity is
instantaneously transmitted through the damper, causing an
immediate displacement of point B by the same quantity $\delta$.
Immediately after the impulse, the B position becomes
$x_\mathrm{B}(t_{i-1}^+) = x_\mathrm{B}(t_{i-1}^-) + \delta$.
This instantaneous response reflects the incompressible nature of
the liquid and the mechanical coupling through the hydrodynamic
interaction. Between two consecutive impulses, the mobile surface
attached to the actuator is immobile ($\dot{x}_\mathrm{C} = 0$),
while the confined fluid and the opposite surface attached to the
cantilever relaxes. Then, the differential equation reduces to
$\lambda(D)\cdot\dot{x}_\mathrm{B}(t) + K\cdot x_\mathrm{B}(t)
= 0$. Along this interval, we approximate $\lambda$ as a constant
equal to its previous value calculated at $t_{i-1}^+$, the
beginning of the relaxation motion:
$\lambda_{i-1} = 6\pi\eta R^2/[x_\mathrm{C}(t_{i-1}^+) -
x_\mathrm{B}(t_{i-1}^+) - b]$. This approximation is justified
provided the $\Delta D$ variation over the time interval $\tau$ is
small compared to $D$, which is typically satisfied in SFA
measurements where displacement steps are kept small. The movement
exhibits an exponential decay with time:
\begin{equation}
  x_\mathrm{B}(t) = x_\mathrm{B}(t_{i-1}^+)
  \exp\!\left[-\frac{K(t - t_{i-1})}{\lambda_{i-1}}\right]
  \label{eq:relax}
\end{equation}
representing the gradual relaxation of the cantilever toward its
new equilibrium position.

The iterative numerical calculation proceeds sequentially through
initialization with $x_\mathrm{B}(0) = 0$ and
$x_\mathrm{C}(0) = D(0)$, followed by a loop over each step $i$
from 1 to $N$. At each step, the C position is updated as
$x_\mathrm{C}(i\tau) = x_\mathrm{C}((i-1)\tau) + \delta$, before
the instantaneous B displacement is applied
$x_\mathrm{B}(i\tau^+) = x_\mathrm{B}(i\tau^-) + \delta$. The
new separation is then calculated,
$D_i = x_\mathrm{C}(i\tau^+) - x_\mathrm{B}(i\tau^+)$, with a
concomitant determination of the damping coefficient
$\lambda_i = 6\pi\eta R^2/(D_i - b)$. Finally, at the end of step
$i$, the cantilever deflection $x_\mathrm{B}(t)$ is
$x_\mathrm{B}((i+1)\tau^-) =
x_\mathrm{B}(i\tau^+)\exp(-K\tau/\lambda_i) =
(x_\mathrm{B}(i-1) + \delta)\exp(-K\tau/\lambda_i)$.

The experimentally measured force corresponds to
$F_i = -K\cdot x_\mathrm{B}(t)$, directly relating the cantilever
deflection to the applied load. The numerical calculation is
terminated when $D_i - b$ approaches the molecular dimensions of
the ionic liquid (typically $\sim$1~nm) to avoid unphysical
divergence of the viscous damping coefficient.

This iterative approach captures the essential nonlinear coupling
that cannot be represented through a simple superposition of
distance steps. Rather, the step-by-step calculation reveals how
the system response depends critically on the instantaneous surface
separation, with shorter distances producing stronger damping and
slower relaxation dynamics. At large separations where $\lambda$
is small (weak damping), the exponential relaxation is rapid and
the cantilever quickly approaches its equilibrium position.
Conversely, at small separations where $\lambda$ becomes large
(strong damping), the relaxation is considerably slower. This
dependency creates a complex feedback mechanism where the force
response to each displacement step influences the damping for
subsequent steps.

The methodology provides a theoretical framework for understanding
force transients often observed in SFA measurements, particularly
when surfaces are approached through discrete positioning steps
rather than continuous motion. These spurious transients were
previously observed in a wide variety of systems, particularly in
complex fluids and in materials comprising nanostructures, such as
those reported in our previous investigations including surfactant
aqueous solutions and lyotropic liquid
crystals~\cite{antelmi1995,antelmi1997,freyssingeas1999,antelmi1999,herrmann2014,herrmann2016},
polyelectrolytes~\cite{kulcsar2004,kulcsar2005,bauer2015}, nanocolloids
and nanocapsules~\cite{atkins1997,spalla1997,kekicheff2013,kekicheff2014}.
The transients appear as overshoots in force measurements that
gradually decay to steady-state values---a behavior that cannot be
captured by static force models or linear dynamic approximations.
The nonlinear damping creates a velocity-dependent apparent
stiffening of the system, where rapid approach rates may result in
measured forces that significantly exceed equilibrium values.

Comparison with experimental data requires careful consideration of
the time scales involved. The relaxation time constant
$\tau_\mathrm{rel} = \lambda/K$ varies dramatically with surface
separation, potentially ranging from milliseconds at large
separations to seconds or minutes at molecular-scale confinement
gaps. This variation has important implications for experimental
protocol design, as insufficient waiting times between successive
displacement steps can lead to cumulative errors in force
measurement. The iterative calculation allows optimization of step
sizes and dwell times to ensure that measurements approach true
equilibrium values while maintaining reasonable experimental
duration for collecting data.

\begin{figure*}[hbtp]
  \centering
  \includegraphics[width=\linewidth]{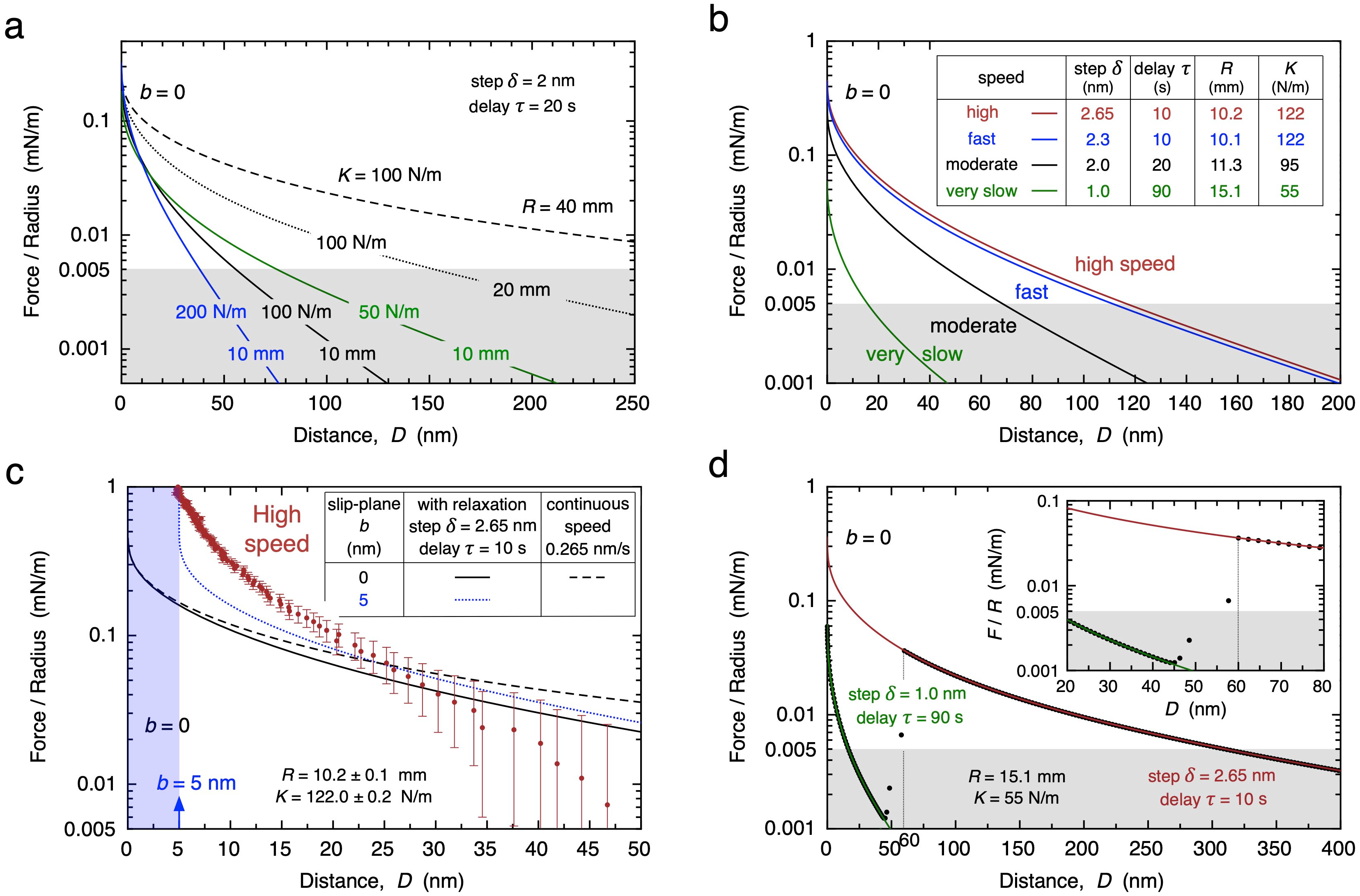}
  \caption{%
    Hydrodynamic force contributions in stepped displacement
    protocols for conventional SFA. All force-distance profiles are
    calculated for a simple liquid having the same viscosity
    $\eta = 37$~mPa$\cdot$s as the IL [C$_2$mim][NTf$_2$]
    investigated by means of conventional SFA (Fig.~1, main text).
    The hydrodynamic forces arise from their partial relaxation upon
    surface movement due to insufficient equilibration time, $\tau$,
    between two consecutive steps when one surface approaches the
    other one by incremental translations $\delta$ (see
    Fig.~\ref{fig:S4}).
    (a) Effect of cantilever spring constant $K$ and radii of
    curvature $R$ of the surfaces in crossed-cylinder geometry: the
    hydrodynamic contribution is reduced if the relaxation time
    (scaling as $R^2/KD$ at a distance $D$) is decreased, that is
    for larger stiffness $K$ and smaller radii $R$. The grey
    shadowy area encompasses the experimentally inaccessible region
    for normalized force measurements ($F/R < 0.005$~mN/m).
    (b) Hydrodynamic contributions to the total measured forces
    reported in Figure~1 (main text) in different protocols
    (``high'', ``fast'', ``moderate'', ``very slow'' speeds). All
    simulations are performed with the same respective experimental
    parameters ($R$, $K$) and with delays, $\tau$, and displacement
    increments, $\delta$, here considered as constant over the full
    surface separation range. The $(\delta,\tau)$ parameters used
    for the numerical calculation correspond to the experimental
    values that were set for measuring the monotonic forces at large
    separations (beyond the oscillatory regime, Fig.~1).
    (c) Hydrodynamic force profiles calculated upon driving the
    surface continuously (dashed black line) or along a stepwise
    approach at the same effective ``high'' speed ($= 0.265$~nm/s)
    with the no-slip plane onset located either at contact ($b = 0$,
    continuous black line) or away ($b = 5$~nm, dotted blue line).
    In all cases the hydrodynamic contributions are smaller than
    the experimental measured forces (red dots) for $D < 30$~nm.
    At larger separations, the uncertainty in force measurements
    calls for caution at this ``high speed''.
    (d) Effect of switching from high velocity (at large separations)
    to very slow speed (at short separations). Despite a factor
    $\sim$25 in velocity reduction, the hydrodynamic forces catch
    up its new profile in only 3 incremental steps, outlining the
    absence of long-term memory effects.
  }
  \label{fig:S5}
\end{figure*}

Figure~\ref{fig:S5} illustrates these numerical predictions for
the hydrodynamic contribution to the measured force in a simple
liquid having the same viscosity $\eta = 37$~mPa$\cdot$s as the IL
(Table~\ref{tab:S1}, refs.\ 1 and 2) investigated here by means of
conventional SFA, namely [C$_2$mim][NTf$_2$].

Experimentally the force sensitivity is increased when surfaces
with larger radii (lower curvatures) are used. This was the
strategy implemented by Gebbie et al.\ who recorded force profiles
for the same IL with $R$ up to 40--50~mm~\cite{gebbie2015}.
Concomitantly the sensitivity is also improved with softer
cantilevers ($K$ in the range 200--50~N/m). However, increasing $R$
enhances the viscous force (as the drag force $F_\mathrm{hydro}$
scales with $R^2$) and decreasing $K$ reduces the restoring force
and increases relaxation times (Fig.~\ref{fig:S5}a). These two
conclusions are clearly outlined by Eq.~\eqref{eq:relax}, stating
that the mechanical equilibrium of the cantilever is achieved at
each step by an exponential relaxation with a characteristic time
scaling as $R^2/KD$. As a consequence of these antagonistic
unfavorable conditions that both increase the hydrodynamic
component, a completely opposite experimental framework must be set:
surfaces with small radii and stiff cantilevers. Because these two
choices go together against force sensitivity in SFA experiments, a
reasonable compromise must be called for. Its search is illustrated
in Figure~\ref{fig:S5}a for different sets ($K$, $R$). There, the
force profiles were calculated using regular intervals $\tau = 20$~s
for bringing the surfaces together by constant displacement
increments $\delta = 2$~nm over the full surface separation range.
Note that this parameter set ($\delta = 2$~nm, $\tau = 20$~s) is
typical of the collected data presented in the main text.

More precisely, Figure~\ref{fig:S5}b reveals the full range of
hydrodynamic responses across the experimental protocols presented
in Figure~1 of the main text. The force profiles are calculated
using the same respective spring constant, $K$, and radius of
curvature, $R$, of the reported experiments (Table of
Fig.~\ref{fig:S5}b). The ``high speed'', ``fast speed'' and
``moderate speed'' protocols were performed with almost similar
$R \sim 10$~mm and $K \sim 100$~N/m, while slower protocols were
carried out with larger radii ($R \sim 15$~mm) and softer
cantilevers ($K = 55$~N/m). Note that all simulations are
calculated with delays, $\tau$, and displacement increments,
$\delta$, that are constant over the full surface separation range.
Their values correspond to the respective experimental ones that
were set for measuring the monotonic forces at large separations
(beyond the oscillatory regime, Figure~1 in main text). The
simulations show that the hydrodynamic force dramatically reduces
both in magnitude and range as long as one slows down the moving
surface and lengthens the equilibration time between successive
positions. The lowest hydrodynamic force profile is obtained when
the moving surface is displaced at ``very low speed'': the magnitude
is below 0.01~mN/m across the entire range, essentially
indistinguishable from zero given the experimental resolution. Not
surprisingly, the highest forces arise at the largest velocity
(``high speed'' profile).

For this ``high speed'' configuration, Figure~\ref{fig:S5}c
compares the measured forces with different simulations using the
same experimental parameters but with different assumptions for
positioning the no-slip plane. Regardless it is set at contact
($b = 0$) or further away ($b = 5$~nm) the experimental forces
emerge significantly above both their experimental error bars and
the calculated hydrodynamic profile for surfaces gap smaller than
30~nm. The observation that hydrodynamic forces are negligible below
this threshold distance for the fastest velocity therefore applies
\textit{a fortiori} to all slower protocols.

The above twofold remark leads to several conclusions. First, even
for relatively fast approach conditions, the choice of adequate
resting times in the stepped displacement procedure allows the
hydrodynamic contributions to be effectively reduced in the
separation regime where monotonic repulsions have been measured.
Secondly, uncertainty in the exact position of the no-slip plane---
which cannot be determined as precisely in conventional SFA as in
dSFA---does not weaken the analysis. Both calculations ($b = 0$
and $b = 5$~nm) predict force profiles with a magnitude well below
the measured forces, suggesting that the monotonic (possibly
exponential) regime might reflect genuine surface interactions rather
than velocity-dependent hydrodynamic effects. Moreover,
Figure~\ref{fig:S5}c raises the question of whether the continuous
motion at average velocity $v_\mathrm{avg} = \Delta\delta/\tau$
would be equivalent to the stepped protocol. The comparison reveals
a striking difference: at equivalent average velocity, a stepwise
displacement consistently produces smaller forces, with differences
being most pronounced at large separations---precisely the distance
range where the zero-force level is determined. For instance, at
$D = 20$~nm, the continuous motion yields $F/R \approx 0.12$~mN/m
compared to $\approx 0.08$~mN/m for the stepped movement. This
occurrence results from the delays that provide discrete relaxation
windows for the system to attain equilibrium through the exponential
factor $\exp[-K\tau/\lambda(D)]$. As long as the delay $\tau$
exceeds several characteristic times $\tau_\mathrm{relax}(D) =
\lambda(D)/K$, the cantilever will have returned close to
equilibrium before the next step occurs. The stepped protocol thus
acts as a low-pass filter, attenuating fast hydrodynamic responses
while preserving slower surface force contributions. For measuring
true equilibrium forces, the stepwise protocols with adequate
equilibration times are not merely being convenient, they are
necessary, since continuous motion systematically overestimates
forces due to unrelaxed hydrodynamic contributions.

Figure~\ref{fig:S5}d illustrates the core of our proposed
methodology allowing the motion to be modified at any surface
separation. Here the force profile is calculated when at large
separations the moving surface traveling at high speed ($v =
0.265$~nm/s) is suddenly slowed down by a factor $\sim$25
($v = 11$~pm/s). In this illustration, the modification of the
effective velocity occurs at $D = 60$~nm. Remarkably, the force
profile catches up its new characteristic response in only 3
incremental steps of displacement. This rapid convergence
establishes that the system responds shortly and does not keep
memory of previous measurements. This behavior is generic of
viscous fluids under confinement: at any separation the force
depends mainly on the current protocol parameters, which validates
the processing of each measurement point as being independent.
Most importantly, with that factual finding in hand, one can
proceed with an analysis of the terms of the procedure and draw
some conclusions and lines of conduct. It appears that extremely
slow surface motions are unnecessary at large separations where
surface forces are weak. For this liquid viscosity, the calculation
shows that faster protocols are sufficient for $D > 60$~nm.
Conversely at smaller separations, the need of accurate measurements
is essential and requires more than just slow movements: long
equilibration times are also needed because nanostructural
rearrangements proceed slowly, requiring the system to fully relax
and to equilibrate at each distance step. Thanks to such a protocol
optimization the total measurement duration can thus be reduced from
12 hours to 4--5 hours without affecting accuracy in critical
regions.

In summary, the numerical simulations led to several findings.
First, it is established that hydrodynamic minimization is achievable
with appropriate protocol parameters. Second, through their periodic
relaxation, the stepwise protocols are inherently superior to
continuous motion. Third, the subtle effects induced by the way the
moving surface is actuated explain the literature discrepancies in
the reported magnitude of the screening lengths. The strong
dependence on measurement conditions---spanning two orders of
magnitude in hydrodynamic contribution---clarifies why different
groups have reported widely varying results. Fast protocols
systematically overestimate the force ranges due to unrelaxed
contributions, creating the illusion of anomalously long-range
interactions. The rapid force profile convergence when one passes
from one protocol to another validates the dependence of measured
forces upon the instantaneous conditions only. This supports the
assumption that hydrodynamic and surface forces contribute to the
total force almost additively. Most importantly, this analysis
confirms the core claim of the main text. Once hydrodynamic
contributions are unambiguously separated and/or minimized through
rigorous experimental protocols---as demonstrated here with both
techniques---long-range forces may arise from non-equilibrium states
in the confined ILs. These apparent measured long-range interactions
are inclined to vanish when the confinement gap thickness is reduced
more slowly with concomitantly longer equilibration times, although
there may remain some additional slow dynamic effects (Figures~1
and~2 of the main text; Figure~\ref{fig:S6}: see
Sec.~5.1). In contrast, at sufficiently slow
velocities with adequate equilibration times, the screening lengths
do converge to values consistent with theoretical predictions for
ionic liquids.

\section{Interplay of hydrodynamic and surface forces contributions}

\subsection{Effect on the structural forces}
\label{sec:structural_forces}

This section examines the effect of velocity and resting time on
the structural forces and their hysteresis between approach and
retraction of surfaces confining ionic liquid films. The measured
force profiles deviate from the expected drainage of a thin liquid
film between solid extended surfaces (Figure~2 (main text) and
Figures~\ref{fig:S1}--\ref{fig:S3},~\ref{fig:S5}), despite the
confirmation that the IL maintains constant bulk viscosity up to the
shear plane and though inertial and acceleration effects are
negligible given low Reynolds numbers ($Re \ll 10^{-4}$). Even
after velocity normalization, small but persistent discrepancies
remain visible in Figure~\ref{fig:S3}. These observations indicate
that the interactions measured at distances preceding the structural
force regime depend on velocity, even in the absence of acceleration
or inertial effects. The standard procedure of subtracting the
calculated hydrodynamic contributions from the total measured forces
to obtain the surface forces assumes that surface velocities are
sufficiently slow for equilibrium conditions to
prevail~\cite{chan1985}. However, the persistence of some
velocity-dependent discrepancies questions not only this assumption
but also the validity of considering surface forces as simple
additive interactions to the velocity-dependent contribution when
interpreting the total measured force, as presented here for
structural forces.

\begin{figure}[hbtp]
  \centering
  \includegraphics[width=0.6\linewidth]{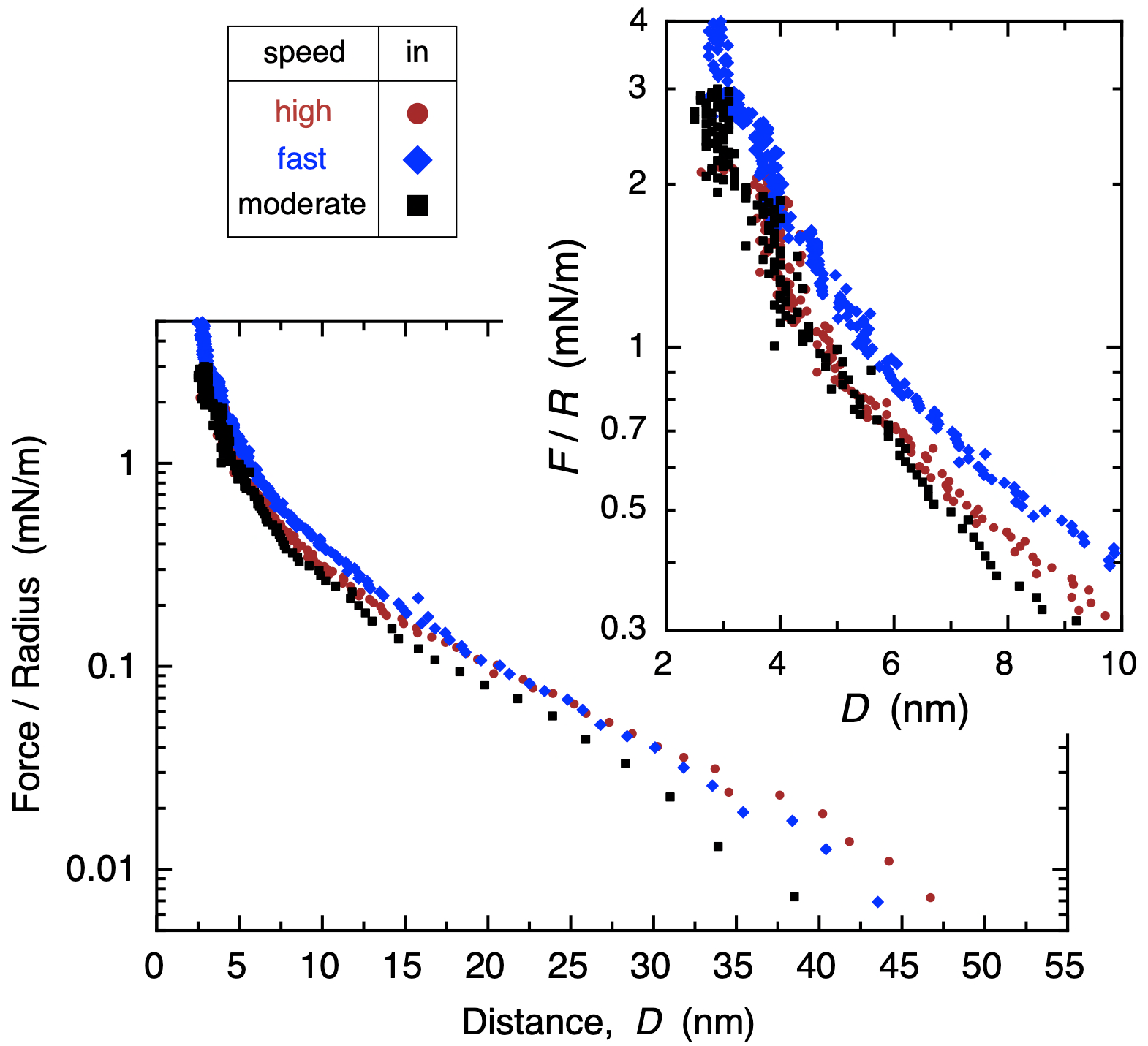}
  \caption{%
    Force-distance profiles for [C$_2$mim][NTf$_2$] confined
    between crossed mica cylinders at different approaching motion
    conditions of the surfaces at high (brown circles), fast (blue
    losanges) and moderate (black squares) speeds (data from Fig.~1,
    main text, with corresponding experimental conditions reported in
    Fig.~\ref{fig:S5}b). The force-distance profiles are composed
    of two distinct regimes with spatial extent and magnitude both
    depending on velocity, stepwise and resting time conditions for
    surface movements (all parameters reported in Sec.~4, SI). At
    large separations the interactions are all monotonic repulsive
    and reversible (see main text). At short separations the
    oscillatory functions of distance are prominently related to the
    structuration of the IL that may not be reversible upon
    decompression. They lie on a repulsive background resulting from
    the interplay of two effects. Faster movements induce larger
    hydrodynamic repulsions, but instead give the confined IL less
    time to undergo structural rearrangements and layering buildup.
    As a result, the compressive energy barrier to squeeze out any
    layer in formation is lowered and the related oscillation in the
    force profile reduces its height or may even disappear. This
    competitive effect is illustrated (inset) by comparing the
    forces collected at the highest velocity (brown circles, high
    speed) that have a smaller magnitude (at small separations) with
    the force collected at a slower velocity (blue squares, fast
    speed).
  }
  \label{fig:S6}
\end{figure}

As illustrated in Figure~\ref{fig:S6}, our observations evidence
a complex IL behavior under confinement and stress, with a subtle
interplay between resting time and solvation (structural) forces
when comparing the force-distance profiles at increasing velocities.
While faster movements generate larger hydrodynamic contributions
causing the total force magnitude to increase at large separations
in the absence of solvation forces, the latter may conversely
prevail at small gaps resulting in higher measured forces despite
slower surface motion (blue losanges, fast speed vs.\ brown circles,
high speed). Faster movements give the confined IL less time to
undergo structural rearrangements and layering buildup. As a result,
the compressive energy barrier to squeeze out any layer in formation
is lowered and the related oscillation in the force profile reduces
its height or may even disappear. Such scenarios, that were already
observed in other fragile systems such as sponge and lamellar
mesophases~\cite{antelmi1995,antelmi1997,freyssingeas1999,antelmi1999}
and in thin
films~\cite{kekicheff2019,kekicheff1994nallet,kulcsar2004,kulcsar2005,bauer2015,kekicheff2013,kekicheff2014,kekicheff2018,kekicheff2019b},
underline the importance of conducting force measurements under
conditions as close as possible to thermodynamic equilibrium.

Another observation is the velocity-dependent extent of the
structured region. Comparing Figure~1 (main text, ``very slow''
speed) with Figure~\ref{fig:S6} reveals that the structural
oscillations begin at $D \sim 8$--9~nm for the slowest approach,
but only below $D \sim 4$--5~nm at higher velocities. This
systematic increase in the thickness of the layered region with
decreasing velocity underscores the need of finite relaxation times
for the IL to build its structure under confinement. At faster
approaches, the confined liquid has insufficient time to develop the
full extent of layering, resulting in a compressed structural regime.
This velocity-dependent structural thickness provides another direct
evidence for the non-equilibrium nature of force measurements at
high approach rates.

The last striking feature is the non-reversibility in the
oscillatory regime upon reversal of the movement. This occurs even
at the slowest possible velocity as illustrated by the semi-log and
linear scales of the force-distance profile that was continuously
measured upon approach of the surfaces and then upon separation
(Fig.~1 bottom, main text). Such observations suggest that at small
gaps the strain field generated by a change in the separation
between two hard walls~\cite{kleman1983} does not induce a pure
elastic response of the confined IL. Three arguments advocate for
this conclusion. The first two remarks are related to the shape of
the oscillations. Although the force profile consists of successive
parts with magnitude and slope increasing as the surface separation
$D$ decreases, the oscillations, though non-linear in shape, are
not parabolae. Secondly, the symmetry of each oscillation is broken
relatively to its minimum. The inside and the outside of the
oscillations appear well differentiated with continuous distortion
of the shape especially upon decompression. Thus, the outward jumps
(for instance, observed at $D \sim 4.1$ and $\sim 5.4$~nm for the
3rd and 4th oscillation respectively) may occur at locations further
away from the expected oscillations minima (that should lie at
grossly mid-distances between two adjacent maxima~\cite{antelmi1995,antelmi1997,freyssingeas1999,antelmi1999,herrmann2014,richetti1995}).
In addition, one has noted that the history of the sample under
study is of importance as both magnitude and shape of the
oscillations may depend on the number of cycles (load/unload)
previously performed at the same contact position. This echoes the
third argument bringing the mechanical instability occurrence
(inward and outward jumps) into scrutiny about their regularity at
a given velocity. Contrary to outward jumps with no regular
positions, the location of the inward jumps as a function of their
rank becomes almost linear upon increasing slowdown: the
periodicity, $d$, approaches a value around 0.8~nm as inferred
from the slope of the line of best fit. This $d$ value is in
agreement with the short-range ordering periodicity reported
previously by AFM~\cite{wang2016} and SFA~\cite{smith2015} and
corresponds to the dimensions of an ion pair. All three sets of
remarks harmonize with respect to claiming that the IL film is
microstructured with bare mica, becomes anisotropic with layering
buildup upon confinement, but does not behave as a smectic phase.
Otherwise, its elastic energy would be lowered by the introduction
or the loss of a layer depending on the sign of the applied external
strain: the relaxation would occur by edge
dislocations~\cite{richetti1995}. Here, the compressive forces
follow a path deviating from a pure elastic parabolic shape, a path
likely to run above the supposed set of intersecting
parabolae~\cite{richetti1995}. On dilation another path is followed,
and even if edge dislocations might exist, their annihilation is not
equivalent to their creation. Actually, as pointed out by de
Gennes~\cite{degennes1972}, some of the defects may bear little
resemblance to dislocations in bulk, as the core region may be
spread along the plane of the layers to such an extent that the
concept of a dislocation line is no longer applicable~\cite{kleman1983,degennes1972,kleman1974,degennes1993,mcgrath1993}.
In the experimental wedge geometry of two crossed cylinders, it may
be energetically less costly to have continuously distorted layers
contouring the curved walls. Whatever the cause, our observations
suggest plastic contributions to the net force and different
activation barriers and delays to squeeze out or incorporate a layer
in the confined medium that involve structural rearrangements and
energetic changes which are propagating from the nucleation site
throughout the thin film, and occur within transit times.

As a result, the structural forces are intrinsically
velocity-dependent---not through an additive velocity-correction
force (the viscosity remains constant as demonstrated in
Sec.~3.2 and viscous contributions are reliably
calculated), but because the force reflects the actual
reorganization state at each measurement timescale. In fact, the
velocity-dependence occurs even if one considers pure elastic
behaviors only. This is for instance the case of confined smectic
mesophases, as first discussed by Pershan and Prost when an external
strain is applied to a smectic confined between two parallel rigid
walls. The relaxation of the induced elastic stress occurs by edge
dislocation loops. Transit times are involved as the nucleation
process follows a Poisson-Boltzmann distribution with a nucleation
rate of stable loops scaling as $\exp(-E_c/k_\mathrm{B}T)$ in
which $E_C$ is the energy barrier of the loop to overcome a critical
pore radius for formation and growth. In a confined medium of
non-uniform thickness, such as a sphere against a flat plane (or
equivalently two crossed cylinders), an array of dislocations
builds~\cite{chan1981}. In the wedge-shaped geometry, the strain
field decays as $\exp(-x^2/\xi z)$ where $\xi$ is the penetration
length and $x$ and $z$ are the coordinates parallel and
perpendicular to the layers, respectively, according to de
Gennes~\cite{degennes1972,degennes1993}. The range of the strain
field is about that of the penetration length, that is a few
reticular distances. For the ILs investigated in this work, $\xi$
would be a few nanometers as the reticular distance is $d \sim
0.8$~nm (see Table~\ref{tab:S1}, Sec.~1 SI, and
refs.~\cite{wang2016,smith2015}). Thus, any change in the
wall-separation induces perturbations not only on the wedge
confinement energy due to the elastic stress of the layered
material, but also on the line tensions associated to each loop,
through the core of the defect itself and the far-field energy from
the distorsion field around the dislocation core~\cite{kleman1983}.
The speed of the moving surface that changes the gap thickness
between the confining walls was shown to be
crucial~\cite{oswald2019,oswald2005,oswald2006}. Some numerical
considerations about the dynamics have also been discussed
previously in the same crossed-cylinder SFA
geometry~\cite{richetti1995}. Yet the concern about the achievement
of thermodynamic equilibrium was emphasized for sound interpretation
of experimental observations, rather than dealing with marked
non-equilibrium effects. The additional complication due to plastic
contributions has also been reported in several systems: a set of
various time relaxations was
evidenced~\cite{herrmann2016,herke1997}, but also additional
slippage events and tapering off to a background drift rate over a
period of tens of seconds with cascades of
instabilities~\cite{herke1997}.

Before concluding this section, a final remark. Note that in the
short-range separation regime, the force profiles are presented
(main text) upon inwards and outwards surface movements only from
large separations to contact or the converse, with no reversed
movement actuated at intermediate separations. This is because the
focus of the work was to investigate the nature and the range of
the long-range monotonic repulsion rather than an investigation of
the solvation structural forces at short separations. Therefore, we
have not sought to recover the inaccessible regions due to the
mechanical instability that causes the surfaces to jump to a new
equilibration separation where the gradient of the force is less
than the cantilever spring constant, $K$. This would have implied
to perform multicycles where the surfaces are brought together,
stopped at specific separations, and then moved outwards to pass the
minimum in force, allowing the position and depth of each
oscillation to be determined, as Perkin's group
reported~\cite{smith2015}. Note also that in all our measurements,
the maximum compressive force $F/R$ remains below 10~mN/m to avoid
artefacts due to the deformation of the glue beneath the mica
sheet~\cite{attard1992,parker1992} (see also Sec.~2)---a value to
be compared to more than 30~mN/m in Perkin's report (Fig.~2 in
ref.~\cite{smith2015}). This difference indicates that we remained
in a regime where repulsive structural forces are still dominant,
but have expelled the outermost few layers only. Therefore, the
force minima in Figure~1 (main text) cannot be compared to the ones
reported for instance in ref.~\cite{smith2015}. Nevertheless, at
very slow speed, the magnitude of $F/R \approx -0.05$~mN/m remains
consistent with the last minimum reported in Figure~2 of
reference~\cite{smith2015}.

In conclusion, structural forces build force-distance profiles that
may be affected by motion conditions of the moving confining
surface. Plastic contributions to the net force, different
activation barriers, delays and transit times to squeeze out or
incorporate layers involving structural rearrangements and energetic
changes, and dependence on confining wall velocity all demonstrate
that the measured structural forces depend strongly on experimental
conditions. Faster approaches yield lower energy barriers
(incomplete layering) while slower approaches allow higher barriers
to be built (more complete organization). This behavior, well
documented in liquid crystals and confined systems, reflects a
fundamental coupling between measurement protocols and structural
states. As a result, caution is advised regarding the usual way of
considering surface forces as simple additive interactions to the
hydrodynamic contribution for interpreting the total measured force,
because other velocity dependence effects are left apart. Only under
quasi-static conditions do the structural forces converge toward
their equilibrium distance profiles.

\subsection{Hydrodynamic vs.\ electrostatic force contributions}
\label{sec:hydro_vs_electro}

The relative importance of hydrodynamic and electrostatic
contributions is examined through a dimensionless analysis
applicable to all constant-velocity measurement protocols. This
framework is relevant not only to the dSFA experiments presented
here, but also to conventional SFA and AFM techniques employing
continuous surface displacement. The dimensionless number $\Pi$
derived below provides a universal criterion for assessing which
force regime dominates under given experimental conditions:
\begin{equation}
  \Pi = \frac{6\pi\eta Rv}{Z\kappa_\mathrm{D}^{-1}
  \exp(-\kappa_\mathrm{D} D)}
  \label{eq:Pi}
\end{equation}

This number is the ratio between the drag viscous force
$F_h = 6\pi\eta R^2 v/D$ and the electrical double layer interaction
$F_e$ acting at large separations, that has a magnitude related to
the surface potential, $\psi_0$, at the onset of the electrical
double layer, and decays exponentially asymptotically with a
screening length $\kappa_\mathrm{D}^{-1}$ as
$F_e = Z\kappa_\mathrm{D}^{-1}\exp(-\kappa_\mathrm{D}D)$
where $Z = 64\pi R\varepsilon_0\varepsilon_\mathrm{r}(k_\mathrm{B}T/e)^2
\tanh^2(e\psi_0/4k_\mathrm{B}T)$. For $\Pi$ larger than one, the
hydrodynamic interaction dominates while the electrostatic force
dominates for $\Pi$ smaller than one.

\begin{figure}[hbtp]
  \centering
  \includegraphics[width=0.6\linewidth]{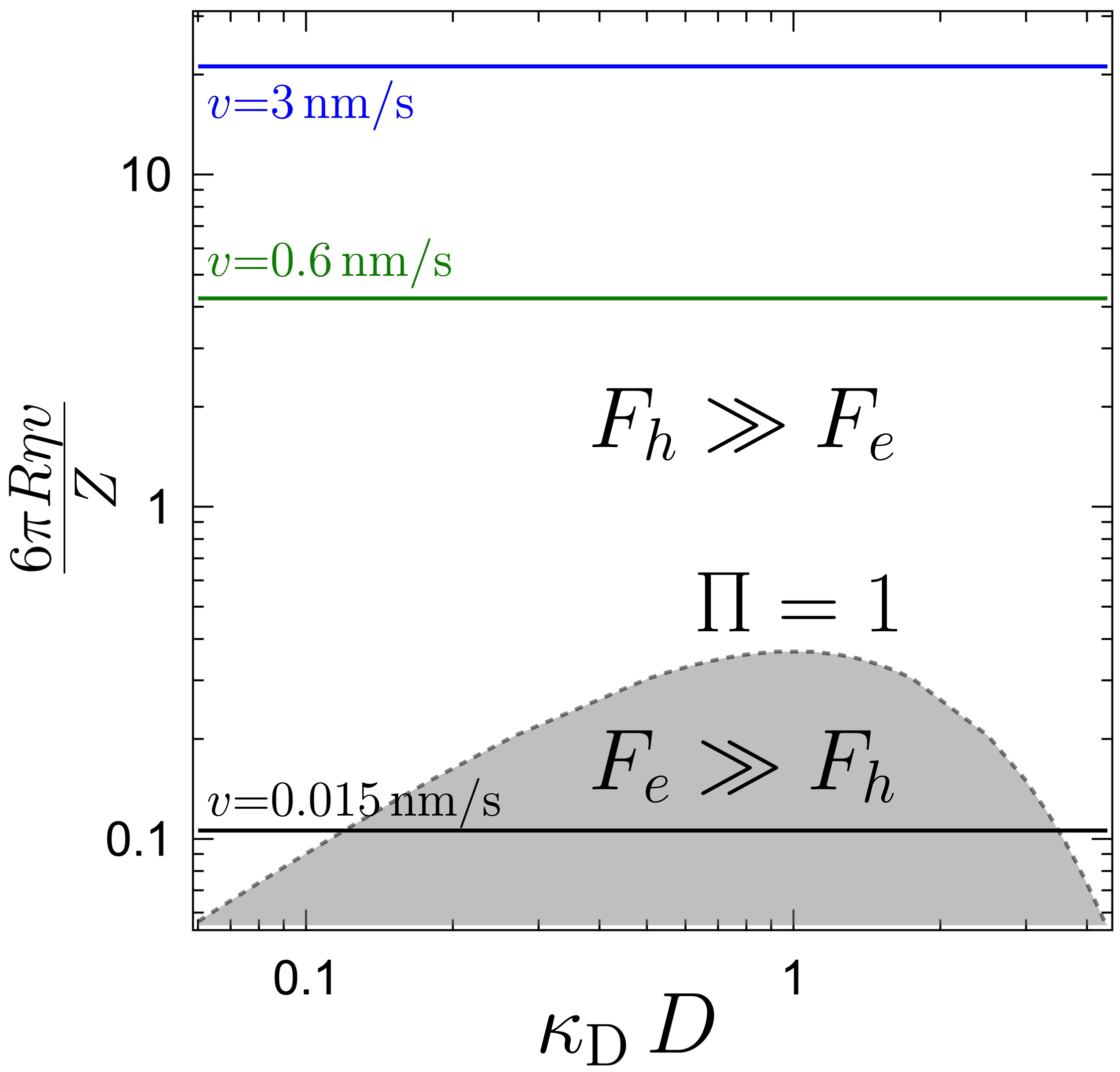}
  \caption{%
    Phase diagram defining force dominance regimes. Dimensionless
    parameter $6\pi\eta Rv/Z$ versus $\kappa_\mathrm{D}D$ shows
    hydrodynamic-dominated (white, $F_h \gg F_e$) and
    electrostatic-dominated (grey, $F_e \gg F_h$) regions separated
    by the $\Pi = 1$ boundary (dashed line). Horizontal lines
    represent experimental velocities of 3.0~nm/s (blue), 0.6~nm/s
    (green), and 0.015~nm/s (black), illustrating the accessibility
    of electrostatic measurements at the lowest velocity.
  }
  \label{fig:S7}
\end{figure}

The equation $\Pi = 1$ is solved numerically for different ratio
values $6\pi\eta Rv/Z$ to find the solutions in terms of
$\kappa_\mathrm{D}D$. The solutions, if they do exist, lead to the
distance range regime where the electrostatic interactions may
dominate the hydrodynamic forces, that is the surface separations
range over which electrostatic interactions can be measured.
Figure~\ref{fig:S7} represents the phase diagram in the coordinates
$6\pi\eta Rv/Z$ versus $\kappa_\mathrm{D}D$ for [C$_4$mim][PF$_6$]
with viscosity $\eta = 157$~mPa$\cdot$s~\cite{yu2012}, dielectric
constant $\varepsilon = 11.4$~\cite{weingartner2008,mizoshiri2010},
and surface potential $|\psi_0| = 16$~mV (see main text). The
dashed line is the representative curve of $\Pi = 1$ in this
coordinate system. If the ratio $6\pi\eta Rv/Z$ is larger than
$\exp(-1) \approx 0.37$, the equation $\Pi = 1$ has no solution and
the hydrodynamic force is always larger, at any distances, than the
electrostatic interaction. For lower values of the ratio, the
equation $\Pi = 1$ has two solutions and electrostatic interactions
are larger than hydrodynamic forces for distances in the interval
given by the two solutions $D_-$ and $D_+$, with $D_- < D_+$. For
distances $D$ larger than $D_+$, the exponential forces are weaker
than the hydrodynamics forces.

The practical measurement of exponential forces superimposed on
hydrodynamic contributions requires careful consideration of the
relative magnitude of the competing interactions. As illustrated in
Figure~\ref{fig:S7}, reliable detection of electrostatic forces
necessitates that $F_e$ exceeds $F_h$ which defines the accessible
measurement window in terms of $\kappa_\mathrm{D}D$. This criterion
ensures that the exponential contribution can be unambiguously
distinguished from the underlying hydrodynamic baseline and that
accurate force decomposition is achievable.

Several important consequences and outputs emerge from this analysis.

First, the assumption of additive force contributions has been
adopted throughout this study. While this assumption appears
reasonable as a starting point, the results from Sec.~3.2
and particularly Sec.~4 indicate that hydrodynamics---or at minimum
the velocity $v$---influences the force profile. This influence,
through a coupling between forces, suggests opportunities for
refining this assumption in future work.

Second, by adopting the force additivity assumption (following
protocols established for dSFA data analysis), behaviors exhibiting
like-exponential characteristics have been identified. At the
slowest velocities, the measurements span two orders of magnitude
in force, and the collected data yield particularly rich insights:
there the exponential nature of the surface force profiles becomes
unambiguous and reliable. At intermediate velocities of 3~nm/s and
0.6~nm/s, some exponential expressions could be inferred despite
noisier data and a more limited force range (spanning only one
order of magnitude; see Figure~2, main text). Under these
conditions, the exponential character becomes less definitive, and
power-law fits would provide comparable quality, highlighting the
inherent flexibility in modeling approaches. Even if successful,
the exponential interpretation proposed by others (see main text)
using fits to large scattered force data and besides over one order
magnitude only is not proven to be physically correct, or that other
models could not be equally successful (and remain equally
unproven!).

This versatility emphasizes the importance of advancing theoretical
frameworks to clarify the fundamental mechanisms governing any
long-range forces at intermediate velocity and to establish more
robust functional forms. It should be noted that this constitutes
an empirical modeling approach designed to describe the experimental
data. Success in fitting does not establish the physical validity
of the model, nor does it preclude other formulations from achieving
comparable results.

Third, an interesting feature of hydrodynamic forces is their
extremely long-range nature. As Figure~\ref{fig:S3} illustrates,
with the force resolution of dSFA, the force remains measurable at
half a micron away from surface contact. Since the technique
determines forces by means of deflections of elastic elements, the
choice of the reference point for defining the zero-force becomes
an important methodological consideration. In previous IL works
reported by other groups, the position of the reference point at
which the zero force is chosen is not specified. As a consequence,
the data analysis can be biased (for instance, setting the zero-force
reference at 100~nm from surface contact implies to subtract the
hydrodynamic force present at this distance from all subsequent
measurements). Conversely, in the current study, such biased
analysis is avoided. Here, the force baseline was established by
averaging the measured force over the interval from 0.8 to
1.0~$\upmu$m, where forces are minimal and stable, thus providing
a robust reference point.

\bibliography{references_SI}
\bibliographystyle{apsrev4-2}